\documentclass[pra,reprint,twocolumn,showpacs,superscriptaddress]{revtex4-1}

\usepackage{graphicx}
\usepackage{dcolumn}
\usepackage{bm}
\usepackage{graphicx}
\usepackage{epsfig}
\usepackage{epsf}
\usepackage{amssymb}
\usepackage{amsmath,empheq,cases}
\usepackage{amsthm}
\usepackage{mathtools}
\usepackage{multirow}
\usepackage[colorlinks=true,linkcolor=blue,citecolor=blue,pdfauthor={ },pdftitle={ },pdfsubject={ },pdfkeywords={ }]{hyperref}
\usepackage{pgfplots}
\usepackage{lipsum}

\usepackage{times}

\theoremstyle{definition}

\usepackage{multirow}
\usepackage{dcolumn}
\newcolumntype{d}[1]{D{.}{.}{#1}}

\begin{document}

\title{Experimental Implementation of Efficient Quantum Pseudorandomness   on a 12-spin System}

\author{Jun Li}
\email{lij3@sustc.edu.cn}
\affiliation{Institute for Quantum Science and Engineering and Department of Physics, Southern University of Science and Technology, Shenzhen 518055, China}
\affiliation{Institute for Quantum Computing, University of Waterloo, Waterloo  N2L 3G1, Ontario, Canada}

\author{Zhihuang Luo}
\affiliation{Beijing Computational Science Research Center, Beijing 100193, China}
\affiliation{Institute for Quantum Computing, University of Waterloo, Waterloo  N2L 3G1, Ontario, Canada}
\affiliation{Institute for Quantum Science and Engineering and Department of Physics, Southern University of Science and Technology, Shenzhen 518055, China}

\author{Tao Xin}
\affiliation{Institute for Quantum Science and Engineering and Department of Physics, Southern University of Science and Technology, Shenzhen 518055, China}

\author{Hengyan Wang}
\affiliation{Department of Physics, Zhejiang University of Science and Technology, Hangzhou 310023, China}

\author{David Kribs}
\affiliation{Department of Mathematics \& Statistics, University of Guelph, Guelph N1G 2W1, Ontario, Canada}
\affiliation{Institute for Quantum Computing, University of Waterloo, Waterloo N2L 3G1, Ontario, Canada} 

\author{Dawei Lu}
\email{ludw@sustc.edu.cn}
\affiliation{Institute for Quantum Science and Engineering and Department of Physics, Southern University of Science and Technology, Shenzhen 518055, China}

\author{Bei Zeng} 
\email{zengb@uoguelph.ca}
\affiliation{Department of Mathematics \& Statistics, University of Guelph, Guelph N1G 2W1, Ontario, Canada}
\affiliation{Institute for Quantum Computing, University of Waterloo, Waterloo N2L 3G1, Ontario, Canada}

\author{Raymond Laflamme} 
\affiliation{Institute for Quantum Computing, University of Waterloo, Waterloo  N2L 3G1, Ontario, Canada} 
\affiliation{Perimeter Institute for Theoretical Physics, Waterloo N2L 2Y5, Ontario, Canada}

\begin{abstract} 
Quantum pseudorandomness, also known as unitary designs, comprise a powerful resource for quantum computation and quantum engineering. While it is known in theory that pseudorandom unitary operators  can be constructed   efficiently, realizing these objects in realistic physical systems can be   a challenging task. In this work, we    study    quantum pseudorandomness generation on a 12-spin nuclear magnetic resonance system. The experimental process is based on the recently proposed   design Hamiltonian approach, which has the merit of   being significantly more efficient than previous protocols.  By applying  random refocusing sequences to the experimental system we    create  a   design Hamiltonian   
the dynamics of   which   quickly  forms   unitary designs. We then use multiple-quantum techniques to measure   spreading of quantum coherences  over  system's degrees of freedom,  and so to probe  the growth of quantum pseudorandomness. The measured multiple-quantum coherence spectra   indicate that   substantial quantum pseudorandomness   have been achieved. 
\end{abstract}


\maketitle

Quantum randomness plays a significant role in quantum information science. It is a fundamentally important resource in   quantum tomography \cite{DCEL09,EB12},       noise characterization \cite{Emerson05, MGE11}, quantum chaos \cite{RY17,CJLY17,KLP18},  quantum metrology \cite{OAGKAL16}, and many other areas.  However, similar to the classical case, the complexity of generating   fully   random transformations on a quantum system  grows  exponentially with the system size  \cite{PZK98}. Therefore, quantum pseudorandomness, often cast as unitary designs more formally, was   put forth as an alternate.
Unitary designs are operationally useful sets of unitaries---a $k$-design   is any ensemble of unitaries capable of simulating up to the $k$-th order statistical moments of the Haar ensemble on average \cite{Dankert09}. Recently, great efforts  have been   devoted  to   identifying efficient constructions of $k$-designs and to exploring their practical uses. In particular,  unitary 2-designs were intensely studied, and were found to have   efficient constructions either exactly from the Clifford group   \cite{DLT02} or approximately from random quantum circuits \cite{Emerson03,Arnaud08,Harrow09,BV10,CHMPS13,brandao2016efficient}.    However,  in experimental  aspects, progress is quite limited   as    unitary designs have been achieved   only in  small-sized physical systems \cite{Emerson07, Chow09, Steffen13, MWOT15}. As the scale of  controllable   quantum  systems  continues to grow rapidly today, realizing pseudorandom   operations on these systems becomes an important and  challenging task.


In this Letter, we study  experimental generation of approximate unitary designs on a 12-qubit spin system, using   techniques of nuclear magnetic resonance (NMR).  On the whole, our study has to   address  two  important problems.  The first problem concerns experimental feasibility. There have been devised a variety of generation protocols that use, e.g.,   polynomial-sized random quantum circuits \cite{Emerson03,Arnaud08,Harrow09,BV10,CHMPS13,brandao2016efficient},   graph state techniques \cite{TM16,MGDM18}, or  random dynamics of design Hamiltonian  \cite{nakata2017efficient}. With feasible experimental realization in mind, we follow the design Hamiltonian approach in our work due to its benefits such as   saving of qubit resources and reducing of time cost, compared with the other protocols. A design Hamiltonian is some random Hamiltonian satisfying that its time-evolutions form unitary designs   spontaneously.  Actually, a concrete form of design Hamiltonian already appeared  in Ref. \cite{nakata2017efficient}, which is  composed of periodically     changing  random spin-glass-type interactions. Here, we show that these disordered interactions can be readily simulated by means of NMR refocusing techniques. Our numerical and   experimental results   indicate that,  evolving the 12-qubit   system under a suitably created design Hamiltonian     is an effective and feasible way of producing pseudorandom evolution operators.  

Our second problem refers to  how to  test    randomness of the  evolution  operators produced in experiment. Recent theoretical studies have suggested that tools such as   out-of-time-order correlators \cite{nakata2017efficient,Yoshida2017}, R\'{e}nyi entanglement entropies \cite{LLZZ18},  or neural networks \cite{AF18} may serve as diagnostics of unitary designs. 
However, there is much lesser experimental study. Such a difficulty can
arise from the complexity in   manipulating and detecting   systems at scale. 
For instance, in Ref. \cite{Yoshida2017} it was shown  that  a natural probe of randomness, namely \emph{frame potential},   can be expressed in terms of   out-of-time-order correlators.
However,   these correlator functions may become difficult to estimate at late times of the     design Hamiltonian evolution, because    they  tend to be   saturated to their corresponding Haar values, which  are exponentially small and can not be determined accurately from experiment. Actually, previous experimental work on the measurement of
out-of-time-order  functions were majorly focused on their
short-time decay part \cite{Garttner17,Jun17}. In our study, we are concerned about not only the short-time, but also the long-time behaviour
of the pseudorandomness generation process, as the former
features the convergence property, and the latter can signal the
onset of pseudorandomness. To this end, we make use of  the multiple-quantum coherence  (MQC) method, a   well-established technique from the realm of solid-state NMR \cite{Baum85,Lacelle91}. Recently, MQCs   attracted great interests for their applications in studying the dynamical and statistical behaviour of complex quantum systems, such as  localization-delocalization transition \cite{AS10,ASK15,WRC18},   buildup of multiparticle entanglement   \cite{GHR18}, and information scrambling \cite{Garttner17}. Here, we show with experimental results that MQC spectra can also be used as a suitable means for detecting the time-development of pseudorandomness   in our 12-qubit system.


\emph{Definitions.}--We start with reviewing the definitions of random unitary  matrices and unitary  designs. Let $\mathbb{U}(d)$ denote the group of $d \times d$ unitary matrices. Consider an ensemble of unitary operators $\mathcal{E} = \left\{ U_i \right\}$ where   $U_i \in \mathbb{U}(d)$. Random unitary matrices $\mathcal{E}_\text{Haar}$ are the ensemble of unitary matrices uniformly distributed with respect to the Haar measure on $\mathbb{U}(d)$. An ensemble   $\mathcal{E}$ is said to be an approximate unitary design if it is close to the Haar ensemble $\mathcal{E}_\text{Haar}$. More precisely,   $\mathcal{E}$ forms an $\epsilon$-approximate $k$-design, if for every monomial $P(U) = U_{i_1 j_1} \cdots U_{i_k j_k} U^*_{m_1n_1}\cdots U^*_{m_kn_k}$ of a degree not more than $k$, its average over     $\mathcal{E}$ is $\epsilon$-close to that over the Haar ensemble $\mathcal{E}_\text{Haar}$, i.e., $\left| (\mathbb{E}_\mathcal{E} - \mathbb{E}_{\mathcal{E}_\text{Haar}})P(U) \right| \le \epsilon$  \cite{brandao2016efficient}.   

Approximate unitary designs can be realized in a number of ways, among which  the design Hamiltonian approach is relatively   easier   to implement  experimentally. A design Hamiltonian is, by definition,   a  physically local   Hamiltonian whose interactions vary randomly at each time step and the dynamics of which   forms a unitary design after a threshold time \cite{nakata2017efficient}.
Put it more strictly, an $\epsilon$-approximate $k$-design Hamiltonian with $l$-local interaction is a random $l$-local Hamiltonian $\mathcal{H}$, where there exists $t_0 > 0$ such that, for most of the time $t \ge t_0$, the propagator $U(t) = \int_0^t\exp(-i \mathcal{H} s)d s$ generated by $\mathcal{H}$ is an $\epsilon$-approximate unitary $k$-design. Here, the shortest such time $t_0$ is called the design time of $\mathcal{H}(t)$.

\begin{figure}[t]
\includegraphics[width=0.95\linewidth]{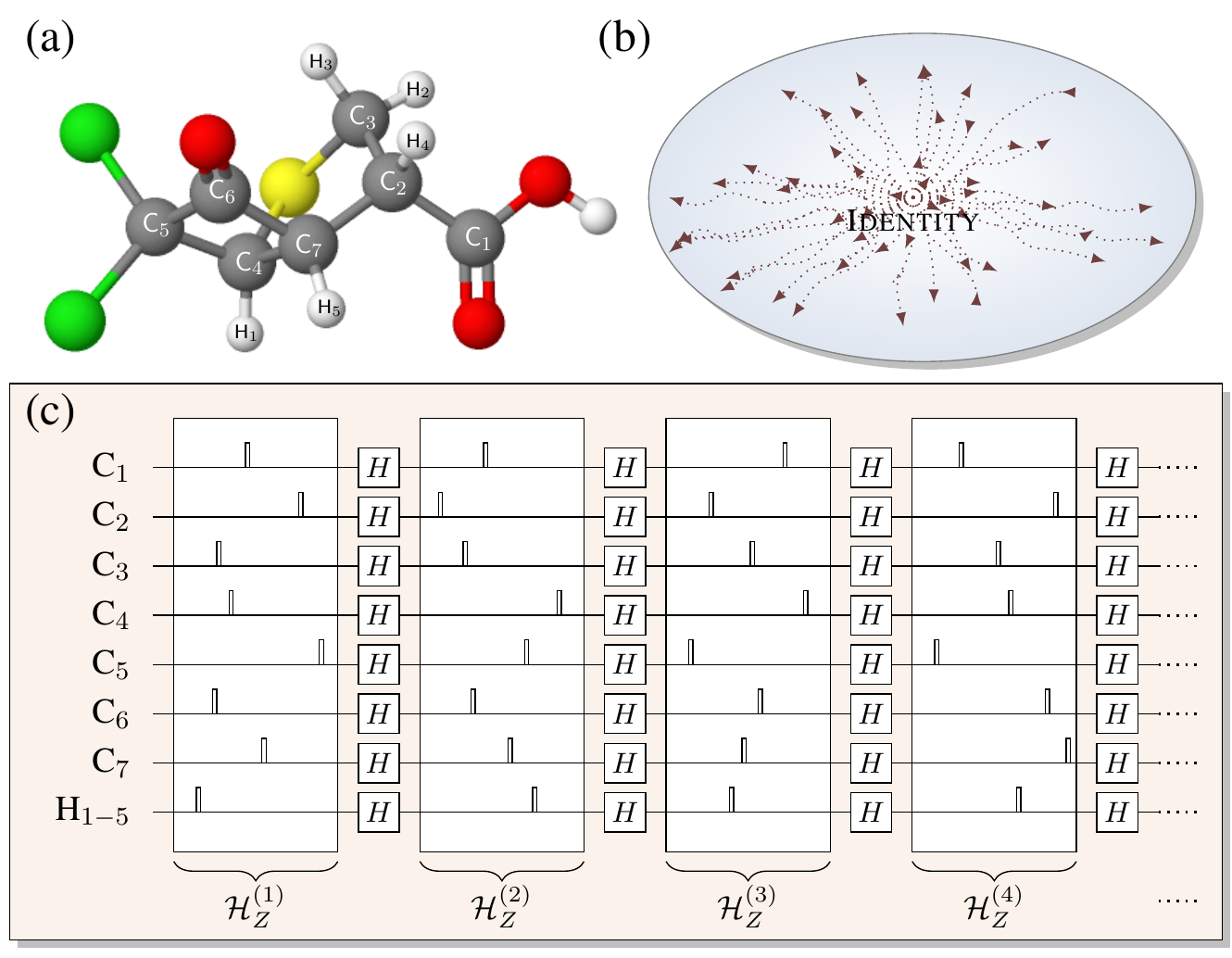}
\caption{(a) Molecular structure of per-$^{13}$C-labeled dichlorocyclobutanone. (b) Intuitive
picture of time-evolution operators generated by a design
Hamiltonian, starting from the identity, and approaching  randomly distributed unitaries over the whole
unitary group as time passes. The trajectories represent different time-evolutions. (c) Schematic illustration of random refocusing pulse sequences that are applied to our 12-qubit system to  produce random Hamiltonian evolutions. The small rectangles represent single-qubit $\pi$ rotations. }
\label{designH}
\end{figure}

\emph{Experimental scheme.}--
In experiment, we chose the per-$^{13}$C-labeled dichlorocyclobutanone derivative dissolved in d6-acetone, which contains 7 labeled carbon nuclei and 5 proton nuclei and hence serves as a 12-qubit system; see Fig. \ref{designH}(a). Experiment was carried on a Bruker Avance III 700 MHz spectrometer at room temperature. The system Hamiltonian under the weak coupling approximation reads 
\begin{equation}
\mathcal{H}_S =   \sum_{i=1}^{12} { \Omega_i \sigma^i_z/2} + \pi\sum_{i<j}^{12} {J_{ij} \sigma^i_z \otimes \sigma^j_z/2},
\end{equation}
where $\Omega_i$ is the precession frequency of the spin $i$ in rotating frame, and $J_{ij}$ is the scalar coupling strength between spins $i$ and $j$; see Supplementary Material \cite{S} for their values.

Our strategy   to achieve quantum pseudorandomness   here   adapts the    design Hamiltonian construction developed in Ref. \cite{nakata2017efficient}. The experimental scheme consists of applying a series of random refocusing pulse sequences  with   change-of-basis operation $H^{\otimes n}$ ($H$ is the Hadamard transform) in between; see Fig. \ref{designH}(c). Refocusing sequence is   commonly used in NMR spectroscopy for adjusting effective couplings between nuclei spins. Usually it is composed of a set of    single-qubit $\pi$ pulses about $x$ or $y$ axis \cite{VC04}. Here, by random refocusing sequence  we mean that    the $\pi$ rotations therein are applied at random time. Such  type of sequences has been previously shown to be useful in constituting randomized dynamical decoupling protocols with good convergence   and stability \cite{SV06,SV08}. Now, to specify the   concrete form of   random refocusing sequence to be used in experiment, we fix the time length of the sequence to be $T/2$ and introduce a set   of 8-tuple column vectors $\lambda = \left\{\lambda^{(m)}: m=1,2,...\right\}$   with their entries  being randomly chosen from the unit interval. For each $m$,  $\lambda^{(m)}$ represents a random refocusing sequence composed of 8 $\pi$ pulses, in which the $i$-th $\pi$ pulse is applied on the $i$-th qubit at time $\lambda^{(m)}_i T/2$.  By applying the $m$-th random refocusing sequence we will get dynamic evolution governed by the following effective disordered Hamiltonian
\begin{equation}
\mathcal{H}^{(m)}_Z   =  \sum_i {\Omega^\text{eff}_{i,m}  \sigma^i_z} + \sum_{i<j} {J^\text{eff}_{ij,m} \sigma^i_z \otimes \sigma^j_z}, 
\label{ZZ}
\end{equation}
where the coefficients $\Omega^\text{eff}_{i,m}$ and   $J^\text{eff}_{ij,m}$ are determined by \cite{S}
\begin{align}
\Omega^\text{eff}_{i,m} & = (1 - 2 \lambda^{(m)}_{i}) \Omega_i,   \label{Omega} \\
J^\text{eff}_{ij,m} & = (1 - 2 \left| \lambda^{(m)}_{i} - \lambda^{(m)}_{j} \right|) J_{ij}.   \label{J}  
\end{align}
Now our design Hamiltonian of the entire sequence  goes: at time $t$, let $m= \lceil t/(T/2) \rceil$, then
\begin{equation}
\mathcal{H}(t) =
\begin{dcases}
\mathcal{H}^{(m)}_Z, &    \text{if $m$ is odd}; \\ 
H^{\otimes n} \mathcal{H}^{(m)}_Z H^{\otimes n}, &    \text{if $m$ is even}.
\end{dcases}
\label{H}
\end{equation} 
Here, in $\mathcal{H}(t)$, the $n$-fold Hadamard transform turns Pauli-$\sigma_z$ bases into   Pauli-$\sigma_x$ bases.  It is expected that the alternate applications of time-evolutions under dual bases  would   quickly approach quantum pseudorandomness.

Note that   our random refocusing sequences realize a design Hamiltonian of the same form as, but with a different parameter set from,      the one proposed in Ref. \cite{nakata2017efficient}.
In the original construction in \cite{nakata2017efficient},     it was theoretically proved  that $\mathcal{H}(t)$ in  Eq. (\ref{H}) can generate an  $\epsilon$-approximate unitary design  within polynomial   time if the  coefficients     are independently  and uniformly distributed. 
Ref. \cite{nakata2017efficient}  also  pointed out that it is   possible to use parameters from different sets, which could result in varied quality and efficiency of   unitary design generation. Here in our construction,   from Eqs. (\ref{Omega})  and (\ref{J}) we have that,   the coefficients $\Omega^\text{eff}_{i,m}$ remain uniformly distributed, but the coefficients $J^\text{eff}_{ij,m}$ are not.  
The main reason that we choose a different parameter set  from the original scheme is due to consideration of   experimental difficulty. Because there exist considerable  decoherence effects   in the sample, it is   desirable that the   pulse  length in experiment be as short as possible. So our construction avoids  coupled evolutions between distant  spins. Besides, since the protons have relatively close resonance frequencies, which  implies longer time required to control them separately, it would be better to perform collective operations on them. With these restrictions in mind, it turns out that, among others,   our experimental sequence is one simplest form of random refocusing sequence that could be realized with reasonable accuracy   on our molecule; see Supplementary Materials for more details \cite{S}. 




\begin{figure}[t]
\includegraphics[width=0.975\linewidth]{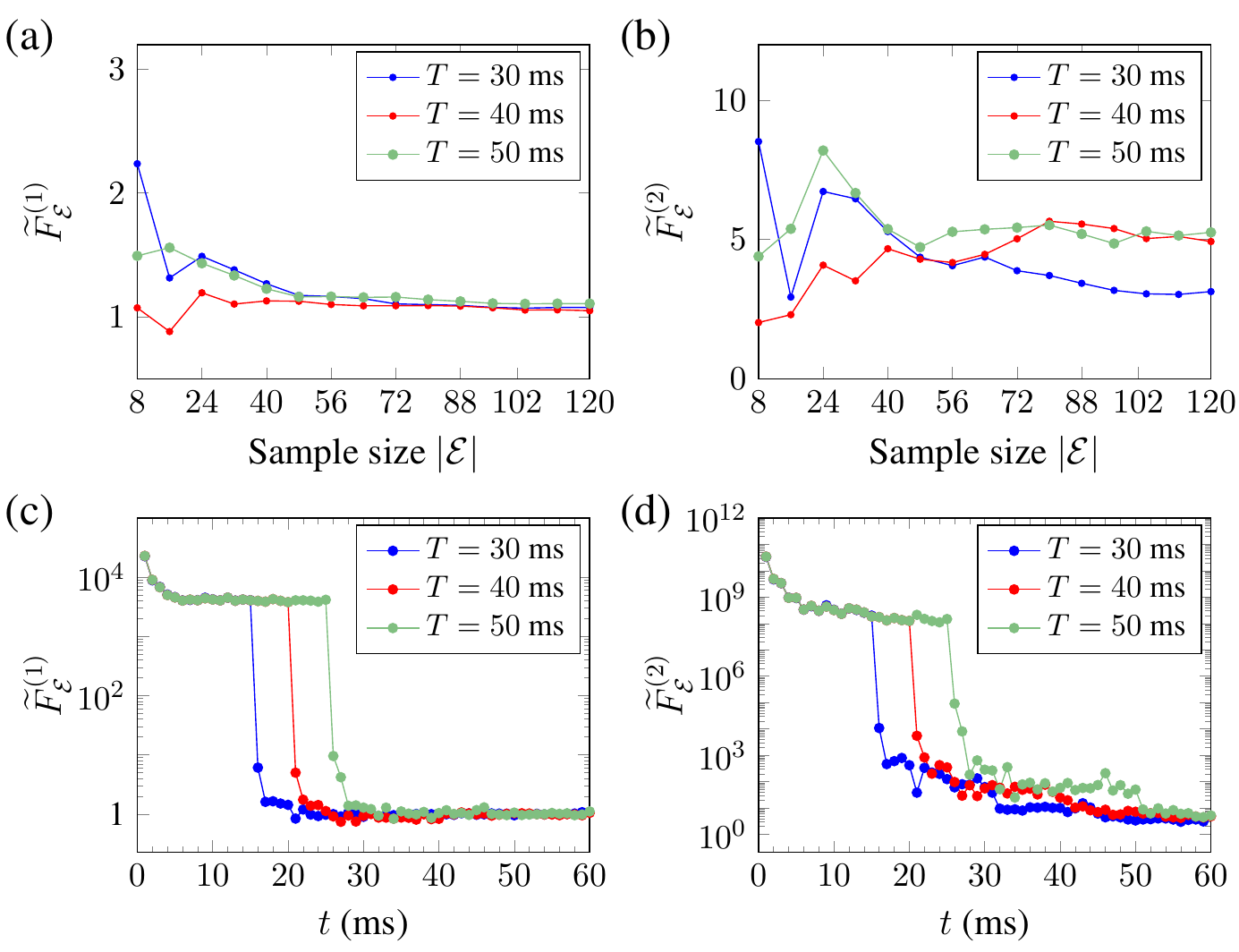}
\caption{Simulation results. (a)-(b) The simulation data here are taken for evolution time of 60 ms. A sampling   of 120  evolutions almost achieves the convergence of $\widetilde F_{\mathcal{E}}^{(k)}$. (c)-(d)   Convergence of first and second frame potential estimated from   sampled   unitary ensembles generated by our design Hamiltonian. $\widetilde F_{\mathcal{E}}^{(k)}$ drops abruptly at the time the change-of-basis operation $H^{\otimes n}$ is applied. As time grows, they eventually approach  the corresponding Haar values.}
\label{simulation}
\end{figure}

We have to check to what extent our Hamiltonian forms an approximate design Hamiltonian.
A  useful test for  unitary designs is made using the notion of  frame potential \cite{RBSC04}. Let $\mathcal{E} = \left\{ U_i\right\}$ be an ensemble,
the $k$-th frame potential is defined as the average of $k$-th powers of the ensemble elements' Hilbert-Schmidt overlaps \cite{Scott2008}  
\begin{equation}
F_{\mathcal{E}}^{(k)} = \frac{1}{\left| \mathcal{E}\right|^2} \sum_{i, j} {\left|\operatorname{Tr}\left( U_i U_j^\dag \right) \right|^{2k}},
\label{FP}
\end{equation}
and there is   $F_\mathcal{E}^{(k)} \ge F_{\mathcal{E}_\text{Haar}}^{(k)} = k!$, where equality holds iff $\mathcal{E}$ is a $k$-design.
Thus the deviation $F_{\mathcal{E}}^{(k)} - F_{ \mathcal{E}_\text{Haar}}^{(k)}$  can serve as a measure of how close $\mathcal{E}$  is to a $k$-design. For large-sized systems $d \ge  k$, $\mathcal{E}$   must contain at least $\left| \mathcal{E}\right| \ge d^{2k}/k!$  unitaries to become a $k$-design \cite{RS09}. This implies that  exact frame potential calculation is intractable.  We thus have to turn to   statistical estimation. Note that
\begin{equation}
F_{\mathcal{E}}^{(k)}  =   \frac{d^{2k}}{\left| \mathcal{E}\right|^2}    + \frac{\left| \mathcal{E}\right|(\left| \mathcal{E}\right|-1)}{\left| \mathcal{E}\right|^2}  \widetilde F_{\mathcal{E}}^{(k)},
\end{equation}
where 
\begin{equation} 
\widetilde F_{\mathcal{E}}^{(k)} = \frac{1}{\left| \mathcal{E}\right|(\left| \mathcal{E}\right|-1)}  \sum_{i\ne j  } {\left|\operatorname{Tr}\left( U_i U_j^\dag \right) \right|^{2k}}.
\end{equation}
So one has that, if $d \ge k$, $\left| \mathcal{E}\right| \ge d^{2k}/k!$ and $\widetilde F_{\mathcal{E}}^{(k)} \approx k!$, then $F_{\mathcal{E}}^{(k)} \approx k!$. In numerical simulation,   we statistically generate a sample of unitaries $\mathcal{E}$   based on our random Hamiltonian evolutions   and observe the convergence of $\widetilde F_{\mathcal{E}}^{(k)}$ with respect to sample size   $\left| \mathcal{E} \right|$.   Fig. \ref{simulation}(c-d) show our numerical results for different periodic time $T$, suggesting that for a range of periods   and after about two rounds of evolution, the estimated frame potentials converge to their corresponding Haar values.
The   simulation results   give strong evidences that our design Hamiltonian can generate ensemble of    unitaries with significant amount of randomness. 


\begin{figure*}
\includegraphics[width=0.925\linewidth]{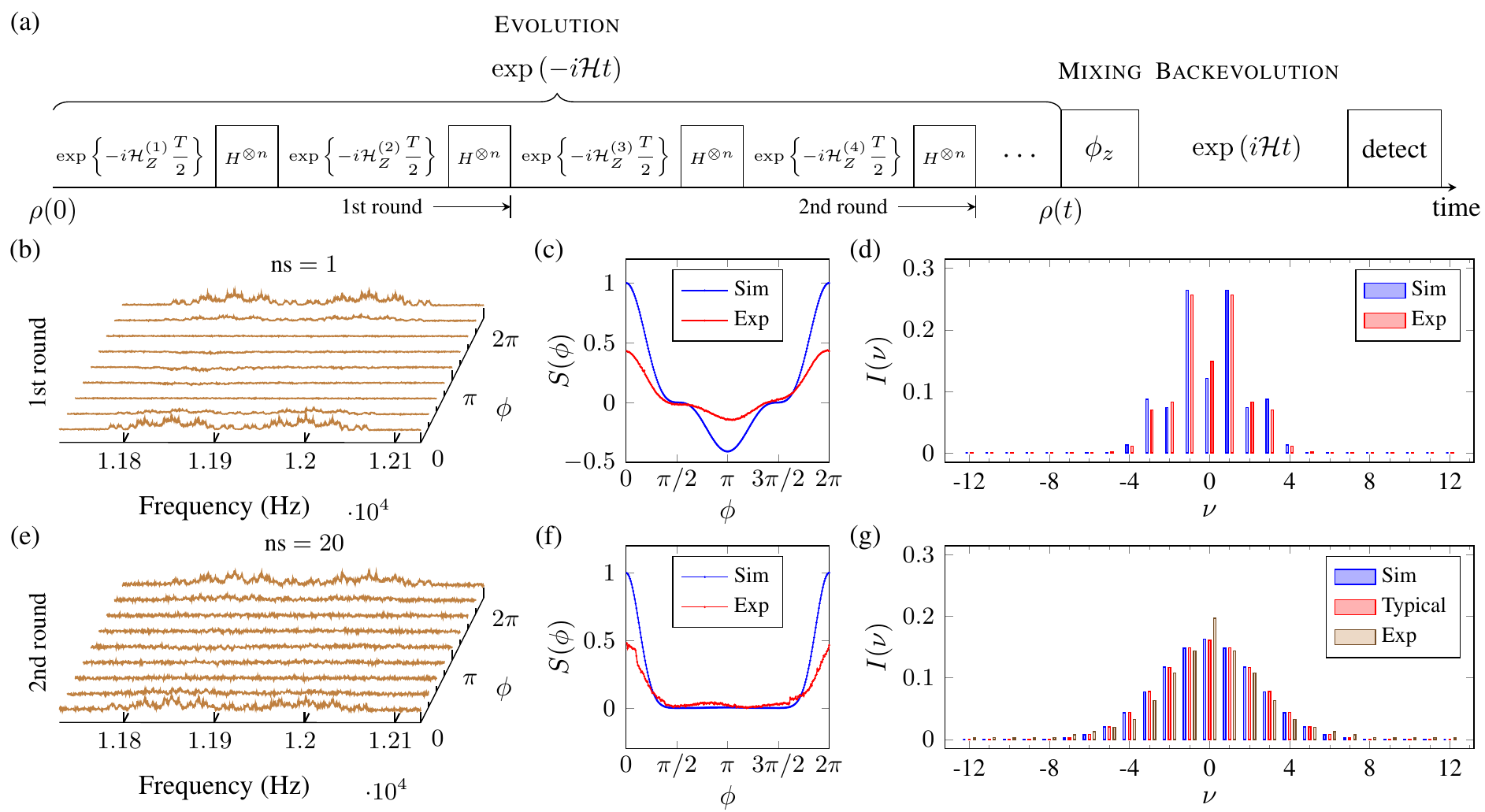}
\caption{(a) Experimental    sequence for  measuring   MQC  spectra. Each round of   design Hamiltonian   evolution and its backward reverse corresponds to an experimental pulse sequence of length 34 ms and 42 ms, respectively.  All   the $\pi$ rotations and   the Hadamard transform take time length 2 ms.  (b)-(g) Experimentally measured MQCs of the evolved state $\rho(t)$ at the first round (b-d) and the second round (e-g) of design Hamiltonian evolution. (b) and (e) A chosen set of experimental $\text{C}_7$  spectra for illustrating purpose. Here, ns means the number of times   we repeat the  data acquisition for compromising signal   loss due to decoherence.  (c) and (f) Multiple-quantum signals observed at varying rotational angles $\phi = \left\{2\pi l/256: l=1,...,256 \right\} $. The left-right asymmetry seen   here could be due to imperfect time reversion of the dynamics. (d) and (g)   MQC spectral intensity profiles. The intensity for each order has been normalized relative to the total spectral intensity.  The profiles demonstrate that the MQCs  generated in experiment rapidly spread over the system's degrees of freedom.}
\label{result}
\end{figure*}

\emph{Probing quantum pseudorandomness.}-- We  perform   MQC  growth experiments to detect the developed quantum pseudorandomness. An outline of the experimental procedure   is shown in  Fig. \ref{result}(a). Basically, the system undergoes a multiple-quantum process consisting of the following steps: (i) start  from a simple operator $\rho(0)$ (e.g., a localized state); (ii) evolve  under   our design Hamiltonian in Eq. (\ref{H}); (iii) a collective rotational operator $\phi_z = e^{-i M_z \phi}$ is applied, here $M_z = \sum_i \sigma_z^i/2$; (iv) the random evolution is reversed. We then measure the overlap of the final state   with   the initial state, resulting in signal
\begin{align}
S(\phi,t)  & = \operatorname{Tr}\left[ e^{i \mathcal{H} t} \phi_z e^{-i \mathcal{H} t} \rho(0) e^{i \mathcal{H} t} \phi_z^\dag e^{-i \mathcal{H} t} \rho(0)\right]  \nonumber \\
& =  \operatorname{Tr}\left[ \phi_z \rho(t) \phi_z^\dag \rho(t)\right]. \nonumber 
\end{align}
Let $\nu$ denote, for the basis $|i\rangle \langle j|$ in the Zeeman representation, the difference between two quantum numbers: $\nu =  \left\langle i | M_z |i \right\rangle - \left\langle j | M_z |j \right\rangle$. Divide $\rho(t)$ into blocks as $\rho(t) = \sum_\nu \rho_\nu$ where $\rho_\nu$ is the submatrix composed of all the order-$\nu$  elements, and note that $\phi_z   \rho_\nu(t)   \phi_z^\dag = e^{-i \nu \phi} \rho_\nu(t) $, there is thus
\begin{equation}
S(\phi,t) = \sum_\nu  e^{-i \nu \phi} I(\nu,t),
\end{equation}
where $I(\nu,t) = \operatorname{Tr}\left[\rho_\nu^2(t) \right]$. 
Now it is clear that the steps taken above  are to ensure that in observing the multiple-quantum signal, all contributions to a given order of coherence are generated with the same phase.
If we measure   $S(\phi,t)$ as a function of $\phi$ at a fixed time $t$ and then perform a Fourier transform with respect to $\phi$, then we are able to extract all the amplitudes $I(\nu,t)$ of     $\rho(t)$, which is often referred to as the MQC spectrum.  Furthermore, with varying the evolution time $t$ we will see the growth of MQCs. 

What would $I (\nu, t)$ look like typically if the evolution operator $U(t)$ is truly random? 
Intuitively, under  a Haar random  operation, all possible
coherences will be excited with equal probability. Then typically the total intensity within a given order $\nu$ is  related simply   to the number of transitions  consistent with that order. In an $n$-spin system, the number of equivalent configurations for a coherence of order $\nu$ is $C_{2n}^{n-\nu}$, which is well approximated by $2^{2n} (n \pi)^{-1/2} \exp(-\nu^2/n)$ for $n >6$. In this picture,   the resulting MQC spectrum  typically    shows   a     Gaussian pattern, i.e.,
$I_{\text{typical}} (\nu) \sim \exp(-\nu^2/n)$.
More   details of the derivation are presented in Supplementary Materials \cite{S}.   This typical behaviour has been observed extensively in solid-state NMR where the spin  dynamics is rather complex \cite{Baum85,Lacelle91}. Accordingly, we expect   in our   experiment that   at long time $t$,
\begin{equation}
I (\nu,t) \to I_{\text{typical}} (\nu).
\end{equation}
Therefore, the essential idea taken here for probing the onset of quantum pseudorandomness is to measure the MQC spectrum, and then compare it with $I_{\text{typical}} (\nu)$.

Fig. \ref{result} shows our experimentally extracted    MQC intensity distributions in our 12-qubit system   at   the first  and   second round  of   design Hamiltonian evolution. Here, the results are taken for   $\rho(0) = \sigma^7_z$, $T= 30$ ms, and a     randomly generated array $\lambda$ whose entries are given in Supplementary Materials \cite{S}.
The shaped pulses for implementing  the  random $\pi$ rotations are obtained from the pulse compiler technique \cite{Ryan08, Jun16}.  The simulated fidelities of these $\pi$ rotations as well as the Hadamard transform are all above 98.5\%, with consideration of control field inhomogeneities. Pulse imperfections and decoherent effects accumulate over   rounds of evolution, and the subsequent
degradation in performance unavoidably  reduces the
signal-to-noise ratio in     multiple-quantum signal observation.  
Importantly,  while these nonideal processes deteriorate the   overall fidelity of the MQC spectra, a  tendency for coherences of higher order to develop with time is clearly evident in the spectra shown in
Fig. \ref{result}(d,g). In particular, we put the typical MQC profile $I_{\text{typical}} (\nu)$ in  Fig. \ref{result}(g) for comparison. And we find that, the experimentally observed redistribution
of spectral intensity into high-order coherences is a
tangible manifestation of the growth of quantum pseudorandomness
during the evolution period.


\emph{Discussions.}--To generate  pseudorandom   operations requires the ability of making extensive control over the system's degrees of freedom. NMR systems are well suited to study pseudorandomness generation process, featuring  
unique control in preparation, manipulation, and detection.  Thus they make excellent testbeds to realize the ideas. Our   approach to the study  of random spin dynamics employs refocusing technique and multiple-quantum NMR technique. Our experimental results demonstrate the
usefulness of the design Hamiltonian method in generating highly complex evolutions. In particular, there is no need to perform  coupled   operations between physically nonadjacent spins and no fine control of time is required.
Because of the wide applicability of  pseudorandom quantum operators,     we expect the techniques   developed and tested here will find broad applications in future quantum information   protocols.

\emph{Acknowledgments}.  
We are grateful to the following funding sources: NSERC (D. K., B. Z. and R. L.); CIFAR (B. Z. and R. L.). J. L., T. X. and D. L. are supported by the  National Natural Science Foundation of China (Grants  No. 11605005, No. 11875159  and No. U1801661),   Science, Technology and Innovation Commission of Shenzhen Municipality (Grants No. ZDSYS20170303165926217 and No. JCYJ20170412152620376),  Guangdong Innovative and Entrepreneurial Research Team Program (Grant No. 2016ZT06D348).


%

\newpage

\begin{center}
\textsc{\Large Supplementary Materials}
\end{center}

\tableofcontents

\section{Experimental System}
Our  quantum pseudorandomness generation experiment is performed on the 12-spin molecule per-$^{13}$C labeled (1S,4S,5S)-7,7-dichloro-6-oxo-2-thiabicyclo[3.2.0]heptane-4-carboxylic acid. The molecular structure is as follows:
\begin{center}
\includegraphics[width=0.6\linewidth]{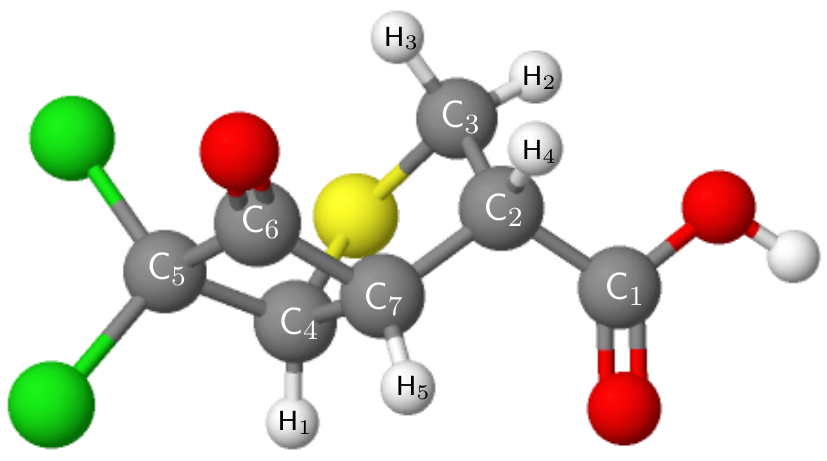}
\end{center}
The following   table gives the   the molecular paramters including $\omega_i$ (diagonal) and $J_{ij}$ (off-diagonal)
\begin{widetext}
\begin{center}
\begin{table}[h]
{\footnotesize\renewcommand{\arraystretch}{1.2}\setlength{\tabcolsep}{5pt}
\begin{tabular}{cD{.}{.}{1}D{.}{.}{1}D{.}{.}{1}D{.}{.}{1}D{.}{.}{1}D{.}{.}{1}D{.}{.}{1}D{.}{.}{1}D{.}{.}{1}D{.}{.}{1}D{.}{.}{1}D{.}{.}{1}}
\hline\hline
  $~~$ & \multicolumn{1}{c}{C$_1$} & \multicolumn{1}{c}{C$_2$} & \multicolumn{1}{c}{C$_3$} & \multicolumn{1}{c}{C$_4$} & \multicolumn{1}{c}{C$_5$} & \multicolumn{1}{c}{C$_6$}  & \multicolumn{1}{c}{C$_7$}  & \multicolumn{1}{c}{H$_1$} & \multicolumn{1}{c}{H$_2$} & \multicolumn{1}{c}{H$_3$} & \multicolumn{1}{c}{H$_4$} & \multicolumn{1}{c}{H$_5$}\\
  \hline
  C$_1$ & 30020.09 &        &         &          &          &          &         &        &   &  &       \\
  C$_2$ & 57.58   & 8780.39  &         &          &          &          &         &        &   &  &       \\
  C$_3$ & -2.00   & 32.67    & 6245.45  &          &          &          &         &        &   &  &        \\
  C$_4$ & 0.02    & 0.30     & 0.00     & 10333.53  &          &          &         &        &   &  &          \\
  C$_5$ & 1.43   & 2.62     & -1.10     & 33.16     & 15745.40  &          &         &        &   &  &          \\
  C$_6$ &  5.54   & -1.66    & 0.00     & -3.53     & 33.16     & 34381.71     &      &        &   &  &           \\
  C$_7$ &  -1.43   & 37.43    & 0.94     & 29.02     & 21.75     & 34.57     & 11928.71 &        &   &   &  &        \\
  H$_1$ &  0.04   & 1.47     & 2.03     & 166.60      & 4.06      & 5.39      & 8.61   & 3307.85 &   &   &  &        \\
  H$_2$ &  4.41   & 1.47     & 146.60   & 2.37      & 0.00      & 0.00      & 0.00     & 0.00    & 2464.15   &   &  &        \\
  H$_3$ &  1.86   & 2.44     & 146.60   & 0.04      & 0.00      & 0.00      & 0.00     & 0.18    & -12.41    & 2155.59  &  &        \\
  H$_4$ &  -10.10 & 133.60   & -6.97    & 6.23      & 0.00      & 5.39      & 3.80     & -0.68   & 1.28     & 6.00     & 2687.69   &        \\
  H$_5$ &  7.10   & -4.86    & 3.14     & 8.14    & 2.36      & 8.52      & 148.50    & 8.46    & -1.00     & -0.36    & 1.30    &  3645.08      \\
\hline
$T_2^*$ (s) & 0.4   & 0.31    & 0.44     & 0.25    & 0.25      & 0.4      & 0.38    & 0.29    & 0.39     & 0.34    & 0.15    &  0.30      \\
\hline\hline
\end{tabular}
}
\end{table}
\end{center}
\end{widetext}
In experiment, the reference frequencies of the $^{13}$C channel and $^1$H channel are set to be $O_1 = 20696$ Hz and $O_2 = 2696$ Hz respectively. So in the rotating frame, the system Hamiltonian takes the form 
\begin{equation}
\mathcal{H}_S =    \sum_{i=1}^{12} {  \Omega_i   \sigma_z^i/2} + \pi\sum_{i<j}^{12} {J_{ij} \sigma_z^i \otimes \sigma_z^j/2},
\end{equation}
where $\Omega_i$ is the precession frequency of the spin $i$, $\Omega_i = -(\omega_i- O_1)$ for $i \le 7$ and $\Omega_i = -(\omega_i- O_2)$ for $i \ge 8$.  

\section{Theory}
A critical problem in our study is to devise an experimental scheme for probing the degree of  pseudorandomness generated from our design Hamiltonian evolution. Three approaches can be identified: (i) estimating frame potential of the generated evolution operators; (ii) measuring a complete set of out-of-time-order correlators (OTOCs); (iii) multiple-quantum coherence (MQC) technique. The first two approaches can quantitatively and completely determine to what extent a unitary ensemble forms a $k$-design, however, they are   practically hard to realize. The last approach probes the spreading   of quantum coherences over system's degrees of freedom. Actually,  it is revealed in Ref. \cite{GHR18} that multiple-quantum coherences is a specific type of OTOCs. Although MQCs measurement does not offer a complete characterization of  pseudorandomness,   it provides rich dynamical and statistical information and is   experimentally    accessible. 

\subsection{Frame Potential}
Frame potential is a quantity measuring the 2-norm distance between the Haar ensemble and the $k$-fold $\mathcal{E}$-channel. For an ensemble of unitary operators $\mathcal{E}$, the $k$-th frame potential is defined by the following   sum
\begin{equation}
F_{\mathcal{E}}^{(k)} = \frac{1}{\left| \mathcal{E}\right|^2} \sum_{i, j} {\left|\operatorname{Tr}\left( U_i U_j^\dag \right) \right|^{2k}}.
\end{equation}
Denote the frame potential for the Haar ensemble as $F_{\mathcal{E}_\text{Haar}}^{(k)}$. Then \cite{RY17}
\begin{enumerate}
\item[(1)]  $F_{\mathcal{E}_\text{Haar}}^{(k)} = k!$, which holds for $k \le d$.
\item[(2)] For any ensemble $\mathcal{E}$,  there is $F_{\mathcal{E}}^{(k)} \ge F_{\mathcal{E}_\text{Haar}}^{(k)}$, here equality holds iff $\mathcal{E}$ is $k$-design.
\end{enumerate}
The method of characterizing a unitary ensemble in terms of frame potential is exact. However, it's rather difficult to estimate $F_{\mathcal{E}}^{(k)}$ in experiment. First, an ensemble has to contain exponential number of elements to become a design \cite{RS09}, it is not realistic to generate exponential number of evolution operators in experiment. Second,   if we make the estimation from a feasible number of evolution operators, measuring overlaps between these evolution operators is difficult. What's more, as the design Hamiltonian dynamics grows sufficiently random, the overlap between two random evolution operators typically gets  exponentially small such that it can not be  determined accurately from experiment.

\subsection{OTOCs}
Let $\mathcal{E} = \left\{U_i \right\}$ be generated from a design Hamiltonian $\mathcal{H}(t)$. The out-of-time-order correlator is defined as 
\[ \langle A U_i B  U^\dag_i A U_i B  U^\dag_i \rangle_\mathcal{E},\]
where $A$ and $B$ are local  observables,   and $\langle \cdots \rangle_\mathcal{E}$ denotes averaging over the ensemble $\mathcal{E}$.

Ref. \cite{RY17} established the following formula
\begin{equation}
F_\mathcal{E}^{(k)} = \frac{d^2}{ d^{2k}}  \sum_{\substack{A_1, ... \\ B_1,...}}  \left|  \sum_{U_i \in \mathcal{E}} \frac{1}{\left| \mathcal{E}\right|} \operatorname{Tr}\left(A_1 U_i B_1 U_i^\dag  \cdots A_k U_i B_k U_i^\dag \right)  \right|^2       \nonumber
\end{equation}
here, summations are over all possible Pauli operators.
This gives that, frame potential can be expressed as a certain average of OTOC functions. Therefore, the effect of design Hamiltonian   evolution on decreasing the frame potential is equivalent to that on the decay of OTOCs. Owing to this close connection, OTOC measurement could thus be used as  another means of pseudorandomness detection.
However, from experimental aspect of view,  it can be readily seen from the above formula that, to get an exact quantification of the random dynamics   an exponential number of OTOCs are involved.  Furthermore, the random dynamics generated from a design Hamiltonian   should quickly saturate the OTOC functions to their Haar random averages, which are exponentially small. Actually, previous OTOC measurement experiments were majorly focused on the short-time decay   rather than the long-time steady behaviour of OTOC dynamics.

\begin{table*}
\begin{center}
{\small\renewcommand{\arraystretch}{1.5}\setlength{\tabcolsep}{3pt}
\begin{tabular}{cccccccccccccc}
\hline\hline
$\nu$ &  0 & $\pm 1$ & $\pm 2$ & $\pm 3$ & $\pm 4$ & $\pm 5$ & $\pm 6$ & $\pm 7$ & $\pm 8$ & $\pm 9$ & $\pm 10$ & $\pm 11$ & $\pm 12$\\
\hline
$I_\text{typical}(\nu)$ & 0.1612 &  0.1488 & 0.1169 & 0.0779 & 0.0438 & 0.0206   & 0.0080 & 0.0025 & 0.0006 & 0.0001 & 1.65e-5 & 1.43e-6 &  5.96e-8  \\
\hline\hline
\end{tabular}
}
\caption{Typical MQC intensity distribution of a random state on a 12-spin system.}
\label{typical}
\end{center}
\end{table*}

\begin{figure}[b] 
\includegraphics[width=0.8\linewidth]{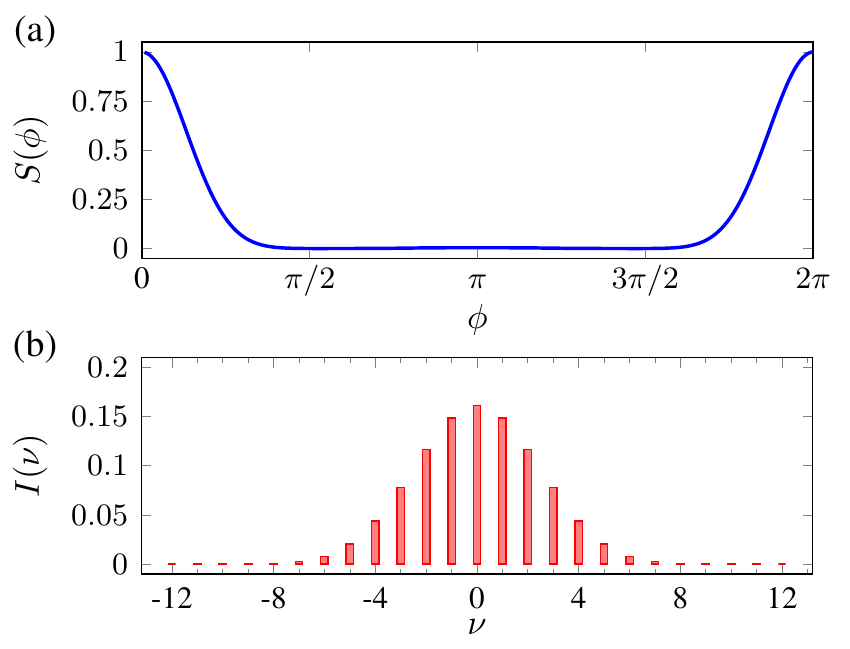}
\caption{Typical MQC distributions when $U$ is random, which shows an approximate Gaussian pattern.}
\label{MQCtypical}
\end{figure} 

\subsection{Statistics of MQC Growth Experiment}
MQC growth experiments were first developed in solid-state NMR.
The basic idea is that, an operator $A(0)$ that is initially localized, e.g., on a single site of a spin network, will evolve
under random evolution $U(t)$ into a vastly more complicated operator $A(t) = U(t) A(0) U^\dag(t)$. The coherences of   $A$ should spread over the entire space. We can perform    MQC growth experiment to get the MQC spectrum  of $A(t)$
\begin{equation}
I_\nu = \operatorname{Tr} \left[A^2_\nu(t) \right],   
\end{equation}
where $-n \le \nu \le n$, and   $A_\nu$ is the submatrix of $A$ composed of all the order-$\nu$ elements. Now the question is, what would $I_\nu$ look like typically if $U(t)$ is random?

The MQC   intensities $I_\nu$ are actually polynomials of elements of $U(t)$. Moments of polynomials on random unitaries can be exactly evaluated [B. Collins, Int. Math. Res. Not. 17, 953 (2003)].
Let $A$ be a traceless and normalized Hermitian operator. If $U$ is an $d \times d$ Haar-distributed unitary matrix, 
and suppose $d$ is large, then
\begin{equation}
\mathbb{E}   \left| \langle \alpha | U A U^\dag | \beta \rangle \right|^2   \approx \frac{\left\| A\right\|^2}{d^2} = \frac{1}{d^2}.
\end{equation}
In a system of $n$ spins, the number of transitions with a given $\nu$ ($-n \le \nu \le n$) is given by a binomial distribution
\begin{equation}
\mathcal{N} (\nu, n) =  \binom{2n}{n-\nu}   = \frac{(2n)!}{(n+\nu)! (n-\nu)!}.  
\end{equation}
The MQC spectrum thus takes the form 
\begin{equation}
I_\text{typical} (\nu ) \approx    \frac{1}{d^2} \binom{2n}{n-\nu}   \sim  \exp\left( - \nu^2/n \right).
\end{equation}
When $n=12$, this gives the distribution shown in Table \ref{typical}.

\section{Experimental Methods}

\subsection{Random $\pi$ Pulse Sequence}

Our experiment of  quantum pseudorandomness generation is based on the design Hamiltonian approach. Our design Hamiltonian is given by Eq. (2-5) of the main text. Such type of design Hamiltonian can be implemented through  the NMR refocusing technique. As we have described in the main text, the sequence that we use is specified by a random  array $\lambda$. 

The principle that a random refocusing   sequence would result  in an effective Hamiltonian of the form    Eq. (2-4) of the main text  can be seen by just considering the   simple 2-qubit case. See the following figure, where the rectangles represent $\pi$ pulses about $x$ (or $y$) axis:
\begin{center}
\begin{tikzpicture}[scale=0.75]

\draw (-2.2,2) rectangle (-1.8,3); 
\draw (2.8,2) rectangle (3.2,3); 
\node [above]  at (-2,3) {$\pi$};
\node [below] at (-2,2) {$\lambda_1 T$};
\node [above]  at (3,3) {$\pi$};
\node [below] at (3,0) {$T$};
\draw (0.6,0) rectangle (1,1);
\draw (2.8,0) rectangle (3.2,1); 
\node [above]  at (0.8,1) {$\pi$};
\node [above]  at (3,1) {$\pi$};
\node [below] at (0.8,0) {$\lambda_2 T$}; 
\draw (-3,0) to (3,0);
\draw (-3,2) to (3,2);
\end{tikzpicture}
\end{center}
Let $\lambda_1, \lambda_2$ be two random numbers, suppose $\lambda_1 \le \lambda_2$, the dynamic evolution can   be written as
\begin{equation}
U_T = 
e^{-i \mathcal{H}_S \lambda_1 T} X_1 e^{-i \mathcal{H}_S (\lambda_2-\lambda_1)T}  X_2 e^{-i \mathcal{H}_S (1-\lambda_2)T} X_{1,2}.  	     \nonumber 
\end{equation}
Note that  
\begin{align}
	X e^{-i Z \alpha} X & = e^{i Z \alpha}, \nonumber \\
	X_1 e^{-i Z_1 Z_2 \alpha} X_1 & = e^{i Z_1 Z_2 \alpha}, \nonumber  \\
	X_2 e^{-i Z_1 Z_2 \alpha} X_2 & = e^{i Z_1 Z_2 \alpha}. \nonumber   
\end{align}
Substituting these formulas into $U_T$, one can get
\begin{equation}
U_T = e^{-i \mathcal{H}_\text{eff} T},	     \nonumber 
\end{equation}
where 
\[ \mathcal{H}_\text{eff}    =  (1 - 2 \lambda_1)  \sigma^1_z + (1 - 2 \lambda_2)  \sigma^2_z + (1 - 2(\lambda_2 - \lambda_1)) J_{12} \sigma^1_z \otimes \sigma^2_z.  \]
For arbitrary $\lambda_1, \lambda_2$, one will then get   Eq. (2-4) of the main text.
In our experiments, the $\pi$ pulses at the end of each random refocusing sequence can actually be absorbed into $H^{\otimes n}$.

\subsection{Pulse Design and Optimization}
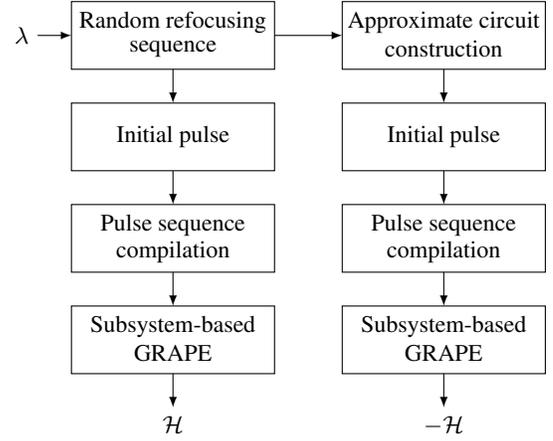
\begin{figure}[t]
\begin{center}
\begin{tikzpicture}[scale=0.9]
 
\draw [-latex] (-2,0) node [left] {$\lambda$} to (-1.5,0);

\draw (-1.5,-0.5) rectangle (1.5,0.5);
\node at (0,0.2) {\small Random refocusing}; \node at (0,-0.2) {\small sequence};
\draw [-latex] (0,-0.5) to (0,-1);
\draw (-1.5,-2) rectangle (1.5,-1);
\node at (0,-1.5) {\small Initial  pulse};
\draw [-latex] (0,-2) to (0,-2.5);
\draw (-1.5,-3.5) rectangle (1.5,-2.5);
\node at (0,-2.8) {\small Pulse sequence}; \node at (0,-3.2) {\small compilation};
\draw [-latex] (0,-3.5) to (0,-4);
\draw (-1.5,-5) rectangle (1.5,-4);
\node at (0,-4.3) {\small Subsystem-based}; \node at (0,-4.7) {\small GRAPE};

\draw [-latex] (1.5,0) to (2.5,0);

\draw (3-0.5,-0.5) rectangle (6-0.5,0.5);
\node at (4.5-0.5,0.2) {\small Approximate circuit}; \node at (4.5-0.5,-0.2) {\small construction};
\draw [-latex] (4.5-0.5,-0.5) to (4.5-0.5,-1);
\draw (3-0.5,-2) rectangle (6-0.5,-1);
\node at (4.5-0.5,-1.5) {\small Initial  pulse};
\draw [-latex] (4.5-0.5,-2) to (4.5-0.5,-2.5);
\draw (3-0.5,-3.5) rectangle (6-0.5,-2.5);
\node at (4.5-0.5,-2.8) {\small Pulse sequence}; \node at (4.5-0.5,-3.2) {\small compilation};
\draw [-latex] (4.5-0.5,-3.5) to (4.5-0.5,-4);
\draw (3-0.5,-5) rectangle (6-0.5,-4);
\node at (4.5-0.5,-4.3) {\small Subsystem-based}; \node at (4.5-0.5,-4.7) {\small GRAPE};

\draw [-latex] (0,-5) to (0,-5.5) node [below] {$\mathcal{H}$};
\draw [-latex] (4,-5) to (4,-5.5) node [below] {$-\mathcal{H}$};
\end{tikzpicture}
\end{center}
\caption{Schematic of pulse design and optimization for design Hamiltonian evolution (left) and its reverse (right).}
\label{pulsedesign}
\end{figure}

Pulse design and optimization for implementing the random $\pi$ sequences and their reverses on a 12-spin system is not an easy task. We have to combine a number of pulse techniques together to achieve the goal.  On the whole, we execute the following step by step:
\begin{enumerate}
\item[(1)] Construct an approximate circuit that   realizes the target evolutions approximately but is as simple as it can be.
\item[(2)] Use selective pulses to implement the single-qubit rotations in the approximate circuit. Here, we  use a pulse sequence compilation program to reduce the phase errors of selective pulse control. The resulting selective pulse sequence serves as a good initial pulse for further gradient-based optimization.
\item[(3)] Use subsystem-based pulse optimization algorithm to further increase the    pulse control fidelity.
\end{enumerate}
The procedure is illustrated in Fig. \ref{pulsedesign}.

\subsubsection{Circuit and Initial Pulse  Construction}
The first step of   pulse optimization  is to construct an initial pulse,   either  from a random  guess or through specific design, which serves as the starting point for subsequent optimization. A suitably constructed initial pulse    makes the optimization procedure easier to reach a final  pulse with satisfying accuracy. According to our practice experiences,  this is especially important  for systems with number of qubits more than seven.

Our strategy for constructing an initial pulse for random $\pi$ sequences and their reverses is to design approximate circuits in terms of an approximate Hamiltonian. Concretely, we consider a simplified coupling network in which we ignore the small couplings and the small differences between large couplings of the original Hamiltonian. Such simplification manifests which couplings
should be majorly accounted for in order to accomplish    reversed    evolutions of $\mathcal{H}_Z^{(m)}$, and thus
enables direct circuit construction. The circuits
thus constructed, if we turn back to the real Hamiltonian,
generate evolutions that deviate the corresponding desired ones   slightly, thus provide   good starting points   for further optimization.

Now we describe the strategy in more details. Consider a refocusing operation 
\[R^{1357}_y(\pi) =  R^1_y(\pi)R^3_y(\pi)R^5_y(\pi)R^7_y(\pi),\]   
notice that
\begin{equation}
U_{-\mathcal{H}_Z^{(m)}} (t) =    R^{1357}_y(\pi)   U_{\mathcal{H}_Z^{(m)}} (t)   R^{1357}_y(\pi)  \cdot U^\text{err}_Z(t)  U^\text{err}_{ZZ}(t),  \nonumber
\end{equation}
where  $U^\text{err}_Z(t)$ and $U^\text{err}_{ZZ}(t)$ are $Z$- and $ZZ$-type evolution errors coming from those evolutions that are not   refocused by $R^{1357}_y(\pi)$, respectively. 
Here, $Z$-type  error terms are easy to handle with. The point is that, whenever there is a rotation about $z$ say $R_z(\gamma)$: (i) if it is
followed by a period of free evolution, their order can be
interchanged; (ii) if it is followed by a transverse rotation
$R_\phi(\theta)$, it can be moved across that rotation according to: $R_\phi(\theta) R_z(\gamma) = R_z(\gamma) R_{\phi-\gamma}(\theta)$. Therefore, $Z$-type errors, whenever encountered, actually need not be executed and can always
be moved one step forward till the end of the circuit \cite{VC04, Ryan08, Jun16}.
The  $ZZ$-type error terms are mainly due to unrefocused couplings $\left\{J_{ij} \right\} / \left\{J_{ij}: \text{only one of $i$, $j$ is in   set $\left\{1,3,5,7 \right\}$} \right\}$. From the parameter table, it is obviously seen that the unrefocused couplings are mostly small, except for $J_{57}$, $J_{2,11}$ and $J_{48}$. Summarizing these observations, we expect that the circuits 
\begin{equation}
R^{1357}_y(\pi)   U_{\mathcal{H}_Z^{(m)}} (t)   R^{1357}_y(\pi)
\end{equation}
are good candidates   on which we seek to achieve   $U_{-\mathcal{H}_Z^{(m)}} (t)$ through pulse optimization.

Another benefit of the above strategy  is that, because we have ignored small couplings, the resulting circuits could be much shorter than those if we do in other ways.

\subsubsection{Selective Pulse Sequence Compilation}
The (approximate) circuit for each $\mathcal{H}_Z^{(m)}$ (or $-\mathcal{H}_Z^{(m)}$) evolution   is composed of free evolutions and 8 $\pi$ rotational gates. To realize the
rotational gates, we  use   frequency selective pulses.
For example, a rotational gate on a specific spin can be realized by
a rotating Gaussian that is on resonance with that spin.
In order that the number of control parameters after
pulse discretization be as few as possible, we adopt relatively large time step length $\tau = 20$ $\mu$s.  

It is important to be aware of that a selective pulse just
approximately implements the target operation. Various
types of errors arise when transferring a circuit directly
into a selective pulse sequence without correction. What's more,
as the number of gates contained in the circuit grows
large, the error accumulation will become increasingly serious. To address this problem, we use the pulse sequence compilation program developed in  Refs. \cite{Ryan08, Jun16}. The compilation
program systematically adjusts the pulse parameters of
an arbitrary input selective pulse sequence so that errors
up to first-order can be corrected. The compilation procedure is efficient. With application of the compilation
method to our pulse sequence, the control accuracy is
greatly improved. Although the compilation program can not eliminate
all control imperfections that higher-order errors still exist, it is still quite useful since that, the pulse sequence after
compilation is of relatively high fidelity and can be used as a good starting point for subsequent gradient-based optimization.

\begin{table*}
\begin{center}
{\scriptsize\renewcommand{\arraystretch}{1}\setlength{\tabcolsep}{6pt}
\begin{tabular}{|cc|cc|cc|cc|cc|}
\hline
\multicolumn{4}{|c|}{\multirow{10}{*}{ $\lambda_\text{rand} = \left(\begin{array}{cccc} 0.3175 & 0.2120 & 0.9879 & 0.4022\\ 0.3164 & 0.0774 & 0.1704 & 0.6207\\ 0.2176 & 0.9138 & 0.2578 & 0.1544\\ 0.2510 & 0.7067 & 0.3968 & 0.3813\\ 0.8929 & 0.5578 & 0.0740 & 0.1611\\ 0.7032 & 0.3134 & 0.6841 & 0.7581\\ 0.5557 & 0.1662 & 0.4024 & 0.8711\\ 0.1844 & 0.6225 & 0.9828 & 0.3508 \end{array}\right)$}} & 
   $\mathtt{ZZYYXYXYZZZZZ}$ & 0.0148 &  $\mathtt{ZIIIXIXXYZXXZ}$ & 0.0069 &  $\mathtt{XXIYZXYYXXZIX}$ & 0.0022  \\ \cline{5-10}
\multicolumn{4}{|c|}{}                   &   $\mathtt{IXIIYXYIZYIZI}$ & 0.0089 &  $\mathtt{YYYIIZZXIYIYY}$ & 0.0013 &  $\mathtt{YXZZZIIYYZYXY}$ & 0.0061  \\ \cline{5-10}
\multicolumn{4}{|c|}{}                   &   $\mathtt{ZXXIXXZIXXYIZ}$ & 0.0122 &  $\mathtt{XIXIZYIYYZYZX}$ & 0.0056 &  $\mathtt{YXXXZXXXYXIZY}$ & 0.0030  \\ \cline{5-10}
\multicolumn{4}{|c|}{}                   &  $\mathtt{ZZXIYXYYZZIIZ}$ & 0.0027 &  $\mathtt{XYIZYYYZZXIYX}$ & 0.0054 &  $\mathtt{ZZXZIIZIYYXXZ}$ & 0.0101 \\ \cline{5-10}
\multicolumn{4}{|c|}{}    &               $\mathtt{IXXXYIXXXIYXI}$ & 0.0091 &  $\mathtt{YIXYZIZIZZZIY}$ & 0.0026 &  $\mathtt{XXIYYIYZXYXIX}$ & 0.0015  \\ \cline{5-10}
\multicolumn{4}{|c|}{}                   &   $\mathtt{ZXZZYXZXYZXXZ}$ & 0.0063 &  $\mathtt{IIXYIYZYYZIYI}$ & 0.0142 &  $\mathtt{XYZZXXYZYZIIX}$ & 0.0141  \\ \cline{5-10}
\multicolumn{4}{|c|}{}                   &    $\mathtt{IZIIIYYIZYIII}$ & 0.0113 &
$\mathtt{YZIIIIZYYYYXY}$ & 0.0126 &  $\mathtt{XIIIZZIXYXZYX}$ & 0.0082  \\ \cline{5-10}
\multicolumn{4}{|c|}{}                   &    $\mathtt{XZYXZYZIIZZZX}$ & 0.0128 &  $\mathtt{YIZZXIXZZYYXY}$ & 0.0064 &  $\mathtt{IZXXXXYZZYXYI}$ & 0.0014  \\ \cline{5-10}
\multicolumn{4}{|c|}{}                   &    $\mathtt{YYIXIZXZIZYYY}$ & 0.0108 &  $\mathtt{ZYIZIXIZIYYYZ}$ & 0.0033 &  $\mathtt{YIZYYIYYZXIIY}$ & 0.0019  \\ \cline{5-10}
\multicolumn{4}{|c|}{}                   &     $\mathtt{YIIIYIZYZXYZY}$ & 0.0122 &  $\mathtt{ZXZIXZXXIZIXZ}$ & 0.0261 &  $\mathtt{YXYYYIYIZXZIY}$ & 0.0059 \\   
\hline
$\mathtt{ZIZZZYYIZXIZZ}$ & 0.0321 &  $\mathtt{XZXYYIXXYXZZX}$ & 0.0052 &  $\mathtt{YXYZYYXXYZYIY}$ & 0.0042 &  
$\mathtt{XXXIZXIXXYYXX}$ & 0.0094  & 
$\mathtt{XYYZYXZIIIIXX}$ & 0.0148 \\ \hline  
$\mathtt{XIYIIYZZYZYYX}$ & 0.0025 & $\mathtt{ZXXIZIXZXYZZZ}$ & 0.0042 &   $\mathtt{ZIXZYYYZYIXXZ}$ & 0.0078   & 
$\mathtt{YYYYXYYIYYXZY}$ & 0.0024 &  $\mathtt{XXXXZYXYZZYIX}$ & 0.0103 \\ \hline 
$\mathtt{YYXIIZIZIYYIY}$ & 0.0081  & $\mathtt{ZIYYIIIIXZYYZ}$ & 0.0060   & $\mathtt{YIZIYIYIZZZZY}$ & 0.0153 &  $\mathtt{XXIYYZIYYIZYX}$ & 0.0119 &  $\mathtt{XZYXZYYYXIYZX}$ & 0.0029 \\ \hline 
$\mathtt{XIXXXZXXXIXIX}$ & 0.0087 &   
$\mathtt{IXXXIXYIZIXII}$ & 0.0009 &  $\mathtt{XIIIXIIYZZIXX}$ & 0.0086 &  $\mathtt{YYXIXIIZXZZIY}$ & 0.0057 &  $\mathtt{YYYYIZIXZZXXY}$ & 0.0097   \\ 
\hline 
$\mathtt{XZZYXZXZIYYIX}$ & 0.0148 &  $\mathtt{ZIXXYIYXIIIZZ}$ & 0.0354 &  $\mathtt{XZYIXIIXYYYXX}$ & 0.0141 &  $\mathtt{IYZZYYYIYXIII}$ & 0.0013 &  $\mathtt{YYXYXIYXZZXIY}$ & 0.0051 \\ \hline  $\mathtt{IZZIXYYYYYXYI}$ & 0.0092 &  $\mathtt{IIZXYZIIXZXXI}$ & 0.0120 &  $\mathtt{YXXIXIYIYZIXY}$ & 0.0027 &  $\mathtt{ZIXYIZZZYXYIZ}$ & 0.0074 &  $\mathtt{ZYYIIXZZYYIYZ}$ & 0.0011 \\ \hline  $\mathtt{ZXZIXYIYYYZZZ}$ & 0.0187 &  $\mathtt{ZXXZIIXYZYXXZ}$ & 0.0124 &  $\mathtt{IZYXIZYZIYXZI}$ & 0.0237 &  $\mathtt{XXXZYIXIYYIIX}$ & 0.0055 &  $\mathtt{ZZYIIZXXZXYYZ}$ & 0.0060 \\ \hline  $\mathtt{ZXYZIYXXXZZIZ}$ & 0.0143 &  $\mathtt{IIZZXYXZXIYZI}$ & 0.0207 &  $\mathtt{ZXYIZZYIXYZZZ}$ & 0.0052 &  $\mathtt{YYXZXIXYYZZZY}$ & 0.0127 &  $\mathtt{IZIIXYYZYXYYI}$ & 0.0156 \\ \hline
$\mathtt{XZXXYZYXYZYYX}$ & 0.0032 &  $\mathtt{IIIIXIZYIIIXI}$ & 0.0261 &  $\mathtt{ZIYXZXXIIIYIZ}$ & 0.0049 &  $\mathtt{IZZXIIYZXXYZI}$ & 0.0056 &  $\mathtt{ZYYXZXIYYXZZZ}$ & 0.0012 \\ \hline  $\mathtt{ZIIIYIXZYXIXZ}$ & 0.0008 &  $\mathtt{ZIXZIYZXIYZYZ}$ & 0.0017 &  $\mathtt{XIZXXIZYYZIZX}$ & 0.0035 &  $\mathtt{ZZYZIXYZZZYYZ}$ & 0.0033 &  $\mathtt{XIZXIYIIYXXXX}$ & 0.0063 \\ \hline  $\mathtt{ZXYIIIXZIYYZZ}$ & 0.0127 &  $\mathtt{XXXYXZZYXYIXX}$ & 0.0213 &  $\mathtt{ZYZZZIYIZYZZZ}$ & 0.0258 &  $\mathtt{ZXIYIIXIYYYXZ}$ & 0.0093 &  $\mathtt{IYZZXYXYZYIZI}$ & 0.0229 \\ \hline  $\mathtt{ZXYZXYZIIIXXZ}$ & 0.0091 &  $\mathtt{XIYIZIYZXIIXX}$ & 0.0136 &  $\mathtt{ZIZYZXZIYZIZZ}$ & 0.0117 &  $\mathtt{YZYZXIXYZXXZY}$ & 0.0061 &  $\mathtt{ZZIZYIXXYZIZZ}$ & 0.0081 \\ \hline
$\mathtt{XXXYYYZXXZYZX}$ & 0.0074 &  $\mathtt{IYYZIYYIZZXZI}$ & 0.0067 &  $\mathtt{IYXXZIXXXZZII}$ & 0.0023 &  $\mathtt{YYXZZYZXYZYZY}$ & 0.0044 &  $\mathtt{IXYZIXZXYZZYI}$ & 0.0052 \\ \hline  $\mathtt{YIYYZYYYXZYIY}$ & 0.0100 &  $\mathtt{XYIZXIIXIYIXX}$ & 0.0037 &  $\mathtt{YZXZYZIZZIXXY}$ & 0.0078 &  $\mathtt{YIYYYXIYXIXXY}$ & 0.0043 &  $\mathtt{YZYZIIXZZXXIY}$ & 0.0122                \\ \hline
\end{tabular}
}

\bigskip

{\scriptsize\renewcommand{\arraystretch}{1}\setlength{\tabcolsep}{6pt}
\begin{tabular}{|cc|cc|cc|cc|cc|}
\hline
\multicolumn{4}{|c|}{\multirow{10}{*}{ $\lambda_\text{rand} = \left(\begin{array}{cccc} 0.4470 & 0.4665 & 0.8616 & 0.8964\\ 0.5876 & 0.4981 & 0.7117 & 0.4822\\ 0.8776 & 0.4874 & 0.8728 & 0.0141\\ 0.4691 & 0.2295 & 0.9380 & 0.6229\\ 0.4374 & 0.0856 & 0.1397 & 0.2311\\ 0.7462 & 0.0674 & 0.3939 & 0.5274\\ 0.4679 & 0.8884 & 0.9806 & 0.7250\\ 0.8608 & 0.2332 & 0.6448 & 0.6074 \end{array}\right)$}} & 
     $\mathtt{IYXIXIYZIIZII}$ & 0.0026 &  $\mathtt{ZZIZIXYYYYXIZ}$ & 0.0089 &  $\mathtt{XIYZIZIZZIYZX}$ & 0.0080  \\ \cline{5-10}
\multicolumn{4}{|c|}{}                   & $\mathtt{XZYZZIXYXZYYX}$ & 0.0021 &  $\mathtt{IXZXZIZIZZYII}$ & 0.0099 &  $\mathtt{XYZIZZXYXIZYX}$ & 0.0089  \\ \cline{5-10}
\multicolumn{4}{|c|}{}                   & $\mathtt{IYXIIXXXXIXYI}$ & 0.0057 &  $\mathtt{ZYXXIIYXYZYYZ}$ & 0.0017 &  $\mathtt{IZYZIXZZXYXXI}$ & 0.0042  \\ \cline{5-10}
\multicolumn{4}{|c|}{}                   &   $\mathtt{YXIXXZYZXXZZY}$ & 0.0096 &  $\mathtt{IZYYZZZYXXZII}$ & 0.0021 &  $\mathtt{IXYIYZIXXZYXI}$ & 0.0074 \\ \cline{5-10}
\multicolumn{4}{|c|}{}    &               $\mathtt{IYZZZIYZIZZII}$ & 0.0094 &  $\mathtt{IXZXXYIIZYXZI}$ & 0.0119 &  $\mathtt{XXYXXYZZIXXZX}$ & 0.0013 \\ \cline{5-10}
\multicolumn{4}{|c|}{}                   &   $\mathtt{IIYXZZZYYZIXI}$ & 0.0038 &  $\mathtt{XZYXIYXIIIXIX}$ & 0.0301 &  $\mathtt{YZXIYZXIIIYXY}$ & 0.0050  \\ \cline{5-10}
\multicolumn{4}{|c|}{}                   &    $\mathtt{YYYYZYZXIIYXY}$ & 0.0299 &  $\mathtt{ZYYYYXXZXIIYZ}$ & 0.0146 &  $\mathtt{ZZIZYXYXYYXIZ}$ & 0.0121  \\ \cline{5-10}
\multicolumn{4}{|c|}{}                   &   $\mathtt{YXZIIZZZZIYZY}$ & 0.0187 &  $\mathtt{YZIZZYXXYIXZY}$ & 0.0120 &  $\mathtt{ZXXXZZYZXXIIZ}$ & 0.0102  \\ \cline{5-10}
\multicolumn{4}{|c|}{}                   &    $\mathtt{IYYZYYIYIIYYI}$ & 0.0259 &  $\mathtt{IXYYIIIXXIYII}$ & 0.0080 &  $\mathtt{ZZZIXZIZZYIXZ}$ & 0.0161  \\ \cline{5-10}
\multicolumn{4}{|c|}{}                   &
$\mathtt{IYXYIYZZXYXZI}$ & 0.0221 &  $\mathtt{IXZIYZYIXXXYI}$ & 0.0083 &  $\mathtt{XZYIYZZYXXYIX}$ & 0.0072 \\  
\hline
$\mathtt{ZZZZZIYYZYXYZ}$ & 0.0045 &  $\mathtt{XYIZIYIXZIZYX}$ & 0.0036 &  $\mathtt{ZIXZIYYYZXXZZ}$ & 0.0032 &  $\mathtt{YXIYIYXXYYZIY}$ & 0.0151 &  $\mathtt{IIZYZIZYYXIZI}$ & 0.0053 \\ \hline  $\mathtt{ZXYZYZIXYIYXZ}$ & 0.0145 &  $\mathtt{XXXZYXIYYIIZX}$ & 0.0099 &  $\mathtt{YIIIZXYXIZXXY}$ & 0.0122 &  $\mathtt{IXXYZYXIXIZYI}$ & 0.0179 &  $\mathtt{YYIZXYYYXIXIY}$ & 0.0018 \\ \hline  $\mathtt{IIZYZYIXZZXZI}$ & 0.0187 &  $\mathtt{XYXYXIXYIXYZX}$ & 0.0018 &  $\mathtt{XXZYIZXXIXXXX}$ & 0.0146 &  $\mathtt{IIZZXXYXXZIZI}$ & 0.0143 &  $\mathtt{XXYYIZYXYZZXX}$ & 0.0078 \\ \hline  $\mathtt{XXIZIXXZIZXXX}$ & 0.0008 &  $\mathtt{ZYIIZYXZZIIYZ}$ & 0.0129 &  $\mathtt{IIIYYYIYZZIXI}$ & 0.0077 &  $\mathtt{IYYXZZZXIYZII}$ & 0.0131 &  $\mathtt{ZZXZZIIIYZYIZ}$ & 0.0128 \\ \hline  $\mathtt{XYXIYYZIXIZXX}$ & 0.0100 &  $\mathtt{ZIZXYZYYIYZYZ}$ & 0.0134 &  $\mathtt{ZZZIIXIXZIXXZ}$ & 0.0197 &  $\mathtt{XZXXZYYZYZXZX}$ & 0.0123 &  $\mathtt{YYZYXIXIXXXZY}$ & 0.0100 \\ \hline  $\mathtt{IYZIYXXYZIYZI}$ & 0.0043 &  $\mathtt{XXXZYXYYXIYYX}$ & 0.0046 &  $\mathtt{XIXIIZXIXZIZX}$ & 0.0075 &  $\mathtt{ZXIZIZIXYIYYZ}$ & 0.0081 &  $\mathtt{XYYXIYIZYZZXX}$ & 0.0154 \\ \hline  $\mathtt{IYZZZYYZXZIZI}$ & 0.0087 &  $\mathtt{YYXIYYZYZYZZY}$ & 0.0148 &  $\mathtt{ZZXZXZXIZZZXZ}$ & 0.0024 &  $\mathtt{YYXZZIYIXXZZY}$ & 0.0157 &  $\mathtt{ZYZYYXYXXXXYZ}$ & 0.0014 \\ \hline  $\mathtt{YIXYXXYZIXIYY}$ & 0.0148 &  $\mathtt{IZIYXZYXXXZXI}$ & 0.0200 &  $\mathtt{XXYYYIXZXYXIX}$ & 0.0034 &  $\mathtt{IZZZIXIXZYYZI}$ & 0.0156 &  $\mathtt{ZZIYYYIXYYZYZ}$ & 0.0048 \\ \hline  $\mathtt{IYZIYXXYIYXZI}$ & 0.0027 &  $\mathtt{XIZIIIIXYZZYX}$ & 0.0264 &  $\mathtt{IYZYXZIIXYYII}$ & 0.0017 &  $\mathtt{IXYXZZXIYYXXI}$ & 0.0010 &  $\mathtt{YZXYIIIZZIZYY}$ & 0.0203 \\ \hline  $\mathtt{YZIXXYIYIYYXY}$ & 0.0145 &  $\mathtt{ZIYXIXXIYYXIZ}$ & 0.0200 &  $\mathtt{XZXYZZZXIZIIX}$ & 0.0158 &  $\mathtt{XZYYZYIZYXXXX}$ & 0.0125 &  $\mathtt{ZIIIXIYZZZYZZ}$ & 0.0101 \\ \hline  $\mathtt{YIZIYIIXIYZXY}$ & 0.0142 &  $\mathtt{YYYXZZZXZIYYY}$ & 0.0120 &  $\mathtt{IXYZIIZXXZZZI}$ & 0.0035 &  $\mathtt{YIXYYZXXZXYZY}$ & 0.0059 &  $\mathtt{IXYIZXXZXXIYI}$ & 0.0086 \\ \hline  $\mathtt{YIYXXYZXZZIZY}$ & 0.0047 &  $\mathtt{IZYYYXXIXYXYI}$ & 0.0131 &  $\mathtt{ZXYZZIXZXZYIZ}$ & 0.0124 &  $\mathtt{XYYXIIXYZIZZX}$ & 0.0145 &  $\mathtt{YIZXIZYIZIIZY}$ & 0.0123 \\ \hline  $\mathtt{ZXYZIYYXYXIXZ}$ & 0.0044 &  $\mathtt{ZIXXIYZZYXXYZ}$ & 0.0133 &  $\mathtt{ZZIXIXZIIYXXZ}$ & 0.0007 &  $\mathtt{ZZYXYIYYYXXIZ}$ & 0.0094 &  $\mathtt{YXXZIZZZXXZXY}$ & 0.0023 \\ \hline  $\mathtt{IYZYYYZZIIIZI}$ & 0.0019 &  $\mathtt{IZYIZZZYYIYYI}$ & 0.0018 &  $\mathtt{ZXYIYIXYXYXIZ}$ & 0.0116 &  $\mathtt{ZYYYYZYIXZIXZ}$ & 0.0034 &  $\mathtt{IZIZIXXXIXYZI}$ & 0.0118                \\ \hline
\end{tabular}
}
\caption{Relative deviations as measured by $\epsilon = \left\|I_\nu(t=2T) - I_\text{Typical} \right\| / \left\|I_\text{Typical} \right\| $ for 100 randomly sampled Pauli basis elements as initial states. Here $\mathtt{I}$, $\mathtt{X}$, $\mathtt{Y}$, and $\mathtt{Z}$ represents the identity operator and the three Pauli operators respectively.}
\label{MQC_longtime}
\end{center}
\end{table*}

\begin{figure*} 
\includegraphics[width=0.85\linewidth]{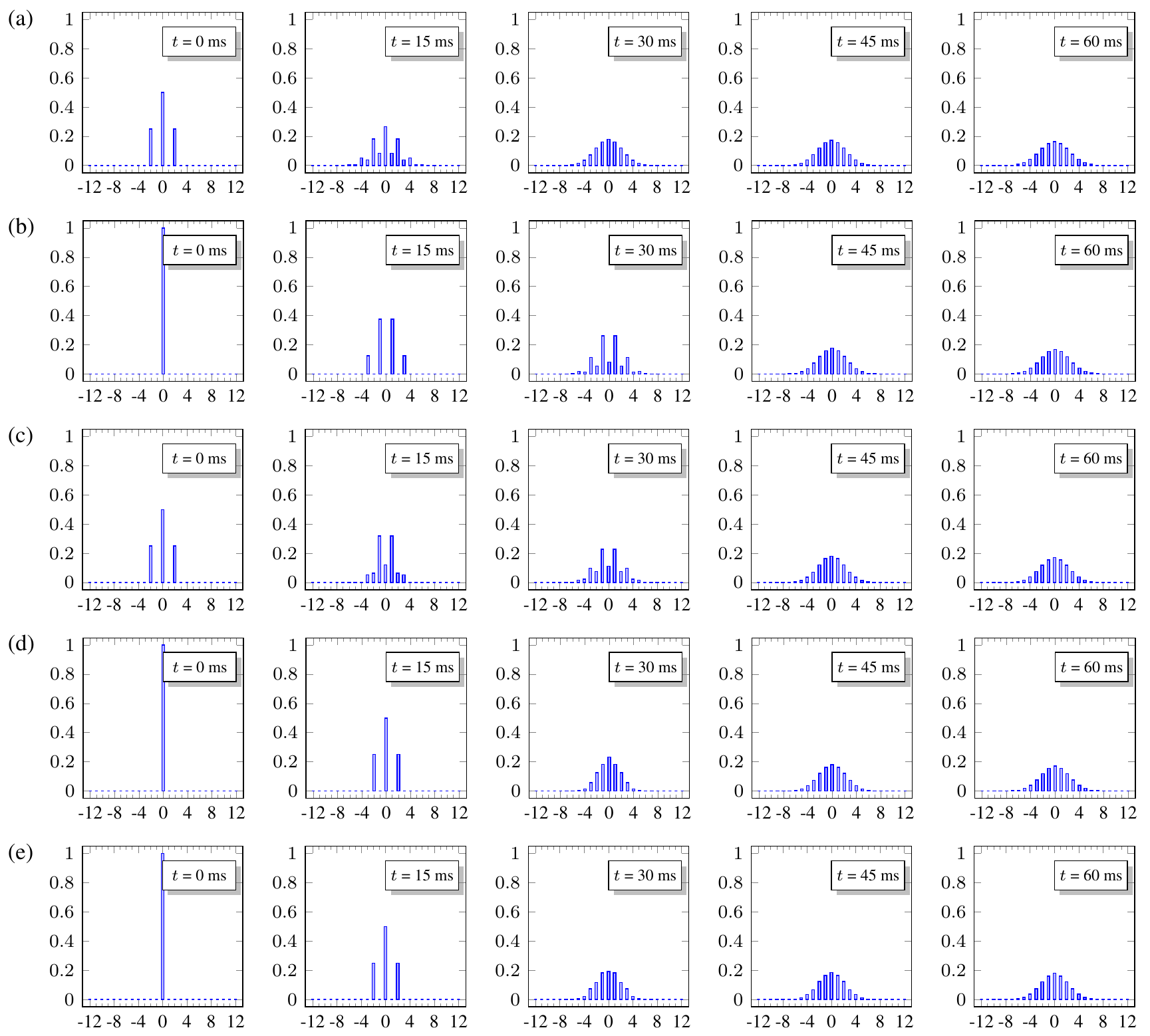}
\caption{Simulated MQC spectra results corresponding to the states  at $t = 0, T/2, T, 3T/2, 2T$ ($T = 30 $ ms), starting from the following initial operators: (a) $\sigma_x^{2}\sigma_x^{7}$; (b) $\sigma_z^{5}\sigma_z^{6}\sigma_z^{10}$; (c) $\sigma_x^{6}\sigma_x^{11}$; (d) $\sigma_z^{9}\sigma_z^{12}$; (e) $\sigma_z^{1}\sigma_z^{7}$. Here, the  refocusing matrix $\lambda$ is randomly generated   as Eq. (\ref{lambda_transient}).} 
\label{MQC_transient}
\end{figure*}

\begin{figure*} 
\includegraphics[width=0.85\linewidth]{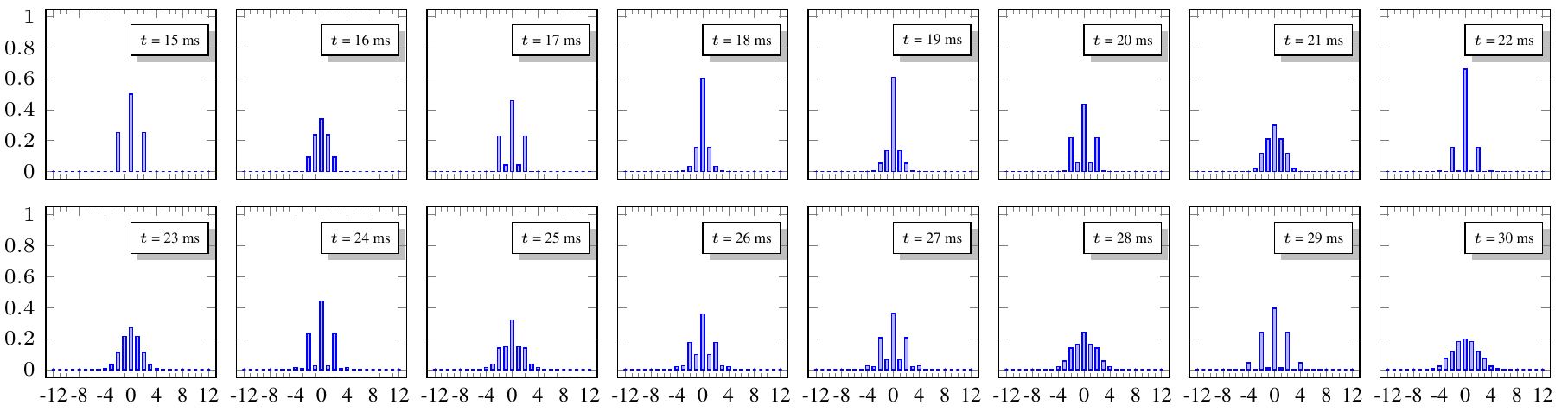}
\caption{Simulated  results showing more detailed transient behaviour of MQC spectra corresponding to the states evolving at $T/2 \le t  \le T$ ($T = 30 $ ms), starting from  $\rho_i = \sigma_z^{1}\sigma_z^{7}$. Here, the  refocusing matrix $\lambda$ is randomly generated   as Eq. (\ref{lambda_transient}).}
\label{Fig3}
\end{figure*}

\begin{figure*} 
\includegraphics[width=0.85\linewidth]{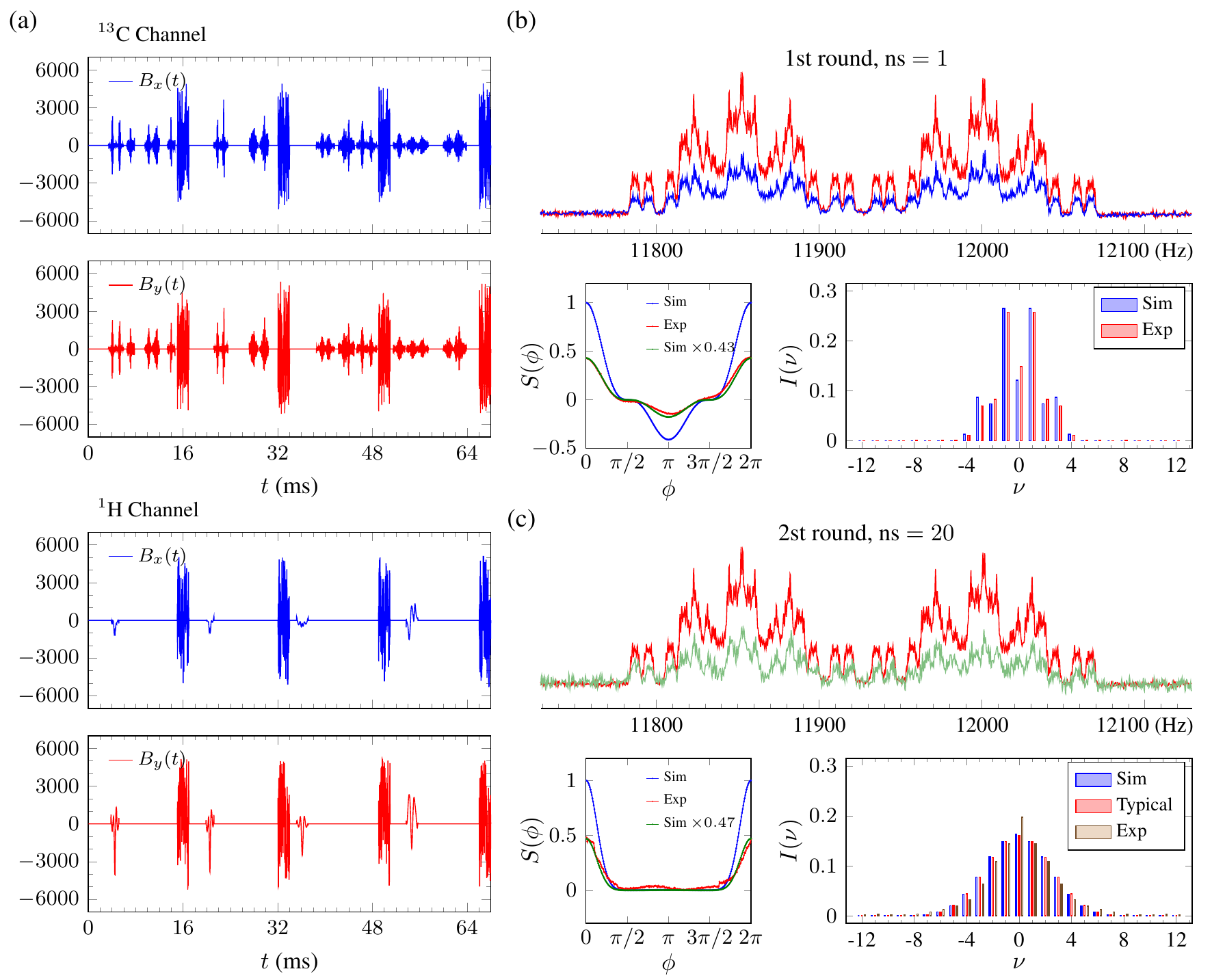}
\caption{Experiment \uppercase\expandafter{\romannumeral1}. (a) Experimental pulse   for generating a random   evolution. (b) and (c) Results of multiple-quantum  signals   and the corresponding MQC spectrum   for the system evolved at 1st round (b) and 2nd round (c). We place the spectrum for initial state $\sigma_z^{7}$    here as a reference spectrum showing the scale of signal-to-noise ratio. The spectrum (color blue) in (b) and the spectrum (color green) in (c) are taken at $\phi = 0$. Ideally, if there are no control errors and no signal loss due to decoherence, they should coincide with the reference spectrum.}
\label{Z7}
\end{figure*}

\subsubsection{Subsystem-based GRAPE}
Gradient ascent pulse engineering (GRAPE) is a numerical algorithm widely used for optimal control pulse search in quantum control. However, it  is challenging to run GRAPE for as large as a 12-qubit system, due to that this involves  computations of $2^{12}$-dimensional matrix multiplications and exponentials that require   substantial amount of memory   and time cost. A variant of GRAPE, namely subsystem-based GRAPE (SSGRAPE) can reduce the computational cost required to some extent \cite{Ryan08}. SSGRAPE works, for our 12-qubit system, as follows. We divide the whole system into two subsystems, $S_A = \left\{ \text{C}_1,\text{C}_2,\text{C}_3,\text{H}_2,\text{H}_3,\text{H}_4\right\}$ and $S_B = \left\{ \text{C}_4,\text{C}_5,\text{C}_6,\text{C}_7,\text{H}_1,\text{H}_5\right\}$, each consisting of 6 spins. The only large couplings between these two subsystems is $J_{\text{C}_2 \text{C}_7}$, so they can be approximately viewed as isolated. Their respective Hamiltonian, $\mathcal{H}_{S_A}$ and $\mathcal{H}_{S_B}$, can be obtained from  $\mathcal{H}_S$ by tracing the other subsystem. Suppose we intend to find a pulse to implement a target operation $U$ (e.g., single-qubit rotation) of the form $U = U_{S_A} \otimes U_{S_B}$. Instead of searching on the whole system, we require the pulse to be optimized should realize desired subsystem operations   on both subsystems. Let $\mathcal{H}_C(t)$ denote the time-dependent   control Hamiltonian, let 
\begin{align}
V(t) & =   \int_0^t \exp\left(-i (\mathcal{H}_S +  \mathcal{H}_C(s) \right), \nonumber \\
V_{S_A}(t) & = \int_0^t \exp\left(-i (\mathcal{H}_{S_A} +  \mathcal{H}_C(s) \right), \nonumber \\
V_{S_B}(t) & = \int_0^t \exp\left(-i (\mathcal{H}_{S_B} +  \mathcal{H}_C(s) \right). \nonumber 
\end{align}
The overall fitness function is $f= \left| \operatorname{Tr} (U ^\dag V (t)) \right|^2$, while
in SSGRAPE we attempt to  maximize the   fitness function
\begin{equation}
f_S = \left( \left| \operatorname{Tr} (U_{S_A}^\dag V_{S_A}(t)) \right|^2 + \left| \operatorname{Tr} (U_{S_B}^\dag V_{S_B}(t)) \right|^2\right)/2,
\end{equation} 
It is expected that when $f_S$ is sufficiently high, then on the whole system, $V(t)$ will get close to the target $U$. In this sense, the 12-qubit GRAPE optimization problem is approximately treated as two 6-qubit optimal pulse control problems.

\section{Some Simulation Results}
In Fig. 2 of the main text, we   showed that, via frame potential estimations, our design Hamiltonian can generate  approximate unitary 2-designs.
Here, before we present our experimental results, we give additional   MQCs simulation results that reveal  more details of  the onset of pseudorandomness.  

To characterize the  relative deviation  between the  MQC spectra   with respect to the typical MQC distribution,
we use the following quantity:
\begin{equation}
	\epsilon(t) =  \frac{\left\|I_\nu(t) - I_\text{Typical} \right\|}{ \left\|I_\text{Typical} \right\|}.
\end{equation}
In Table. \ref{MQC_longtime}, we made calculations on $\epsilon(t=2T)$ for MQCs at the second round of design Hamiltonian evolution,  where the input initial operators $\rho_i$ are randomly selected from the Pauli group, and we have   randomly created two different refocusing matrices. It is found  that, for the tested examples, the relative distance $\epsilon$ rarely exceeds 3\%, which clearly indicates that under our design Hamiltonian evolution, the long-time  MQC spectrum gets close to   the typical distribution.

To see the MQC transient behaviour, we made further computations on the MQC dynamic evolution, which are  shown in Fig. \ref{MQC_transient} and Fig. \ref{Fig3}. The refocusing matrix is randomly generated as follows
\begin{equation}
\lambda = \left(\begin{array}{cccc} 0.9494 & 0.0635 & 0.8321 & 0.4605\\ 0.2564 & 0.3735 & 0.7538 & 0.6455\\ 0.9899 & 0.1663 & 0.6219 & 0.5135\\ 0.3498 & 0.2313 & 0.3941 & 0.8144\\ 0.2085 & 0.0522 & 0.3593 & 0.0972\\ 0.6658 & 0.9018 & 0.0889 & 0.4637\\ 0.9733 & 0.7933 & 0.3417 & 0.5898\\ 0.6227 & 0.373 & 0.5487 & 0.1872 \end{array}\right).
\label{lambda_transient}
\end{equation}
Here, we randomly selected  a set of  local Pauli basis elements as initial states, so that we can see the spreading of coherences. The results are in  Fig. \ref{MQC_transient}. Notice that our design Hamiltonian $H(t)$ (Eq. (2-5) of the main text) is composed of  $H_Z$  and $H_X$, and that $H_Z$  does not change the MQC distribution,  i.e., $H_Z$ just affects  the distribution of the density elements within the same order coherence subspace, so actually we only detect time varying MQC spectra during   the $H_X$ evolutions. Understanding this, we display in Fig. \ref{Fig3} the detailed MQC evolution  during the second stage of the first round  $T/2 \le t < T$. 
To summarize,  the MQC growth experiments  can be used   as a means for     detecting system's coherence distribution, whose temporal and long-time limiting behaviour are in accordance with that the degree of pseudorandomness   grows under the design Hamiltonian evolution.

\begin{figure*} 
\includegraphics[width=0.85\linewidth]{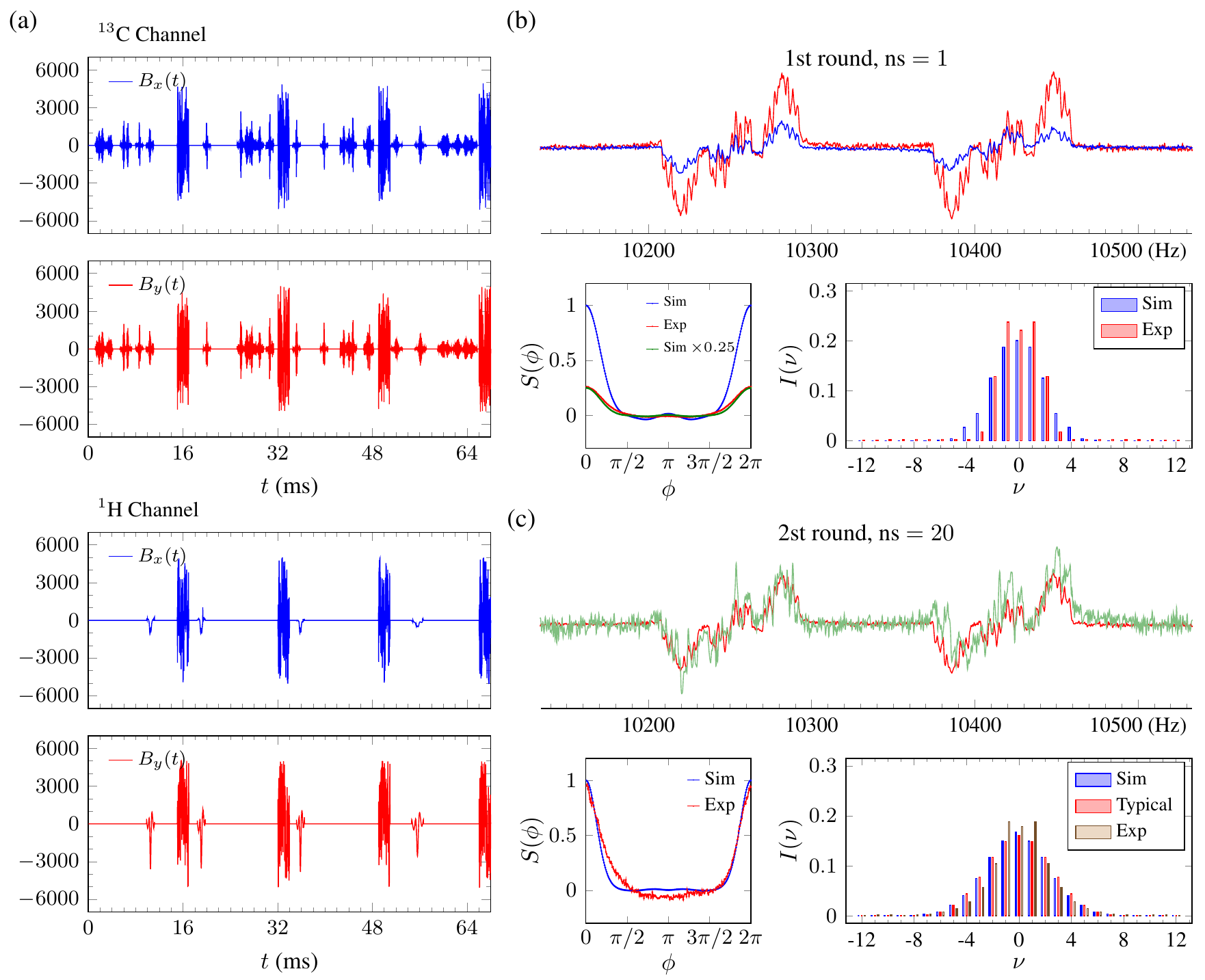}
\caption{Experiment \uppercase\expandafter{\romannumeral2}. (a) Experimental pulse   for generating a random   evolution. (b) and (c) Results of multiple-quantum signals   and the corresponding MQC spectrum   for the system evolved at 1st round (b) and 2nd round (c). We place the spectrum for initial state $\sigma_x^{4}\sigma_z^{5}$    here as a reference spectrum showing the scale of signal-to-noise ratio. The spectrum (color blue) in (b) and the spectrum (color green) in (c) are taken at $\phi = 0$. Ideally, if there are no control errors and no signal loss due to decoherence, they should coincide with the reference spectrum.}
\label{X4Z5}
\end{figure*}

\section{Experimental Results}
We have performed two  experiments observing MQC growth under  evolution by random refocusing sequences   with different parameter sets to demonstrate the process of generating quantum pseudorandomness  through design Hamiltonian evolution. The MQC growth experiment measures the following multiple-quantum signal
\begin{equation}
S(\phi,t)  = \operatorname{Tr}\left[ e^{i \mathcal{H} t} \phi_z e^{-i \mathcal{H} t} \rho(0) e^{i \mathcal{H} t} \phi_z^\dag e^{-i \mathcal{H} t} \rho(0)\right]. 
\end{equation}
Here, the design Hamiltonian $\mathcal{H}$ is specified by $\lambda$.
The experimental results are presented as follows.

\subsection{Experiment \uppercase\expandafter{\romannumeral1}}
Initial state $\rho_0 = \sigma_z^{7}$. The state is prepared  through first destroying  all polarizations except that of H$_5$ and then applying a SWAP gate $\text{SWAP}_{\text{C}_7\text{H}_5}$ ($8 $ms, 0.9883) to transfer the polarization to C$_7$
\[ \rho_{eq} \to \sigma_z^{12}  \xrightarrow{\text{SWAP}_{\text{C}_7\text{H}_5}}  \sigma_z^{7}.  \]
The matrix $\lambda$   is randomly generated given as below
\[
\lambda = 
\left( 
\begin{array}{cccc}
0.2710 &    0.3219 &    0.8206 &  0.3628 \\
0.7585 &	0.7204 &	0.6633 &	  0.7545   \\
0.6796 &	0.8401 &	0.6154 &  0.9943   \\
0.7590 &	0.7336 &	0.6464 &	  0.7608  \\
   0.9323 &  0.3947 &  0.9191  &  0.2523  \\
   0.3520  & 0.3888 &  0.3666  &  0.7918  \\
   0.4772  & 0.8511 &  0.4400  &  0.6621  \\
   0.2982  & 0.2354 &  0.1435  &  0.2457  \\
\end{array} 
\right).
\]
The corresponding experimental pulses for implementing   1st round (34 ms, 0.9910) and 2nd round (68 ms, 0.9642) design Hamiltonian evolution are shown in Fig. \ref{Z7}(a).
Fig. \ref{Z7}(b-c) show the experimental results.


\subsection{Experiment \uppercase\expandafter{\romannumeral2}}
Initial state $\rho_0 = \sigma_x^{4}\sigma_z^{5}$.  The state is prepared through first destroying  all polarizations except that of H$_1$, then applying a SWAP gate $\text{SWAP}_{\text{C}_4\text{H}_1}$ (10 ms, 0.9872) to transfer the polarization to C$_4$, and finally evolving the state under   $J_{45}$
\[ \rho_{eq} \to \sigma_z^{8}  \xrightarrow{\text{SWAP}_{\text{C}_4\text{H}_1}} \sigma_z^{4} \xrightarrow{R_{x}^4(\pi/2)} -\sigma_y^{4} \xrightarrow{J_{45}} \sigma_x^{4}\sigma_z^{5}.  \]
The matrix $\lambda$   is randomly generated given as below
\[
\lambda = \left( 
\begin{array}{cccc}
0.3992 &    0.9113 &    0.7843 &    0.6434 \\
0.1547 &	0.6535 &	0.6132 &	0.8538   \\
0.6927 &	0.7988 &	0.7138 &	0.9148   \\
0.2291 &	0.6864 &	0.6698 &	0.7613  \\
   0.5762 &  0.2014 &  0.3898  &  0.7146  \\
   0.4427  & 0.5866 &  0.9014  & 0.3416 \\
   0.1212  & 0.7092 &  0.8785  &  0.6002  \\
   0.6999  & 0.1389&  0.1223  &  0.3085  \\
\end{array} 
\right).
\]
The corresponding experimental pulses for implementing   1st round (34 ms, 0.9894) and 2nd round (68 ms, 0.9448) design Hamiltonian evolution are shown in Fig. \ref{X4Z5}(a).
Fig. \ref{X4Z5}(b-c) show the experimental results.

%
%
%
%
%
%


\begin{thebibliography}{45}%
\makeatletter
\providecommand \@ifxundefined [1]{%
 \@ifx{#1\undefined}
}%
\providecommand \@ifnum [1]{%
 \ifnum #1\expandafter \@firstoftwo
 \else \expandafter \@secondoftwo
 \fi
}%
\providecommand \@ifx [1]{%
 \ifx #1\expandafter \@firstoftwo
 \else \expandafter \@secondoftwo
 \fi
}%
\providecommand \natexlab [1]{#1}%
\providecommand \enquote  [1]{``#1''}%
\providecommand \bibnamefont  [1]{#1}%
\providecommand \bibfnamefont [1]{#1}%
\providecommand \citenamefont [1]{#1}%
\providecommand \href@noop [0]{\@secondoftwo}%
\providecommand \href [0]{\begingroup \@sanitize@url \@href}%
\providecommand \@href[1]{\@@startlink{#1}\@@href}%
\providecommand \@@href[1]{\endgroup#1\@@endlink}%
\providecommand \@sanitize@url [0]{\catcode `\\12\catcode `\$12\catcode
  `\&12\catcode `\#12\catcode `\^12\catcode `\_12\catcode `\%12\relax}%
\providecommand \@@startlink[1]{}%
\providecommand \@@endlink[0]{}%
\providecommand \url  [0]{\begingroup\@sanitize@url \@url }%
\providecommand \@url [1]{\endgroup\@href {#1}{\urlprefix }}%
\providecommand \urlprefix  [0]{URL }%
\providecommand \Eprint [0]{\href }%
\providecommand \doibase [0]{http://dx.doi.org/}%
\providecommand \selectlanguage [0]{\@gobble}%
\providecommand \bibinfo  [0]{\@secondoftwo}%
\providecommand \bibfield  [0]{\@secondoftwo}%
\providecommand \translation [1]{[#1]}%
\providecommand \BibitemOpen [0]{}%
\providecommand \bibitemStop [0]{}%
\providecommand \bibitemNoStop [0]{.\EOS\space}%
\providecommand \EOS [0]{\spacefactor3000\relax}%
\providecommand \BibitemShut  [1]{\csname bibitem#1\endcsname}%
\let\auto@bib@innerbib\@empty
\bibitem [{\citenamefont {Dankert}\ \emph
  {et~al.}(2009{\natexlab{a}})\citenamefont {Dankert}, \citenamefont {Cleve},
  \citenamefont {Emerson},\ and\ \citenamefont {Livine}}]{DCEL09}%
  \BibitemOpen
  \bibfield  {author} {\bibinfo {author} {\bibfnamefont {C.}~\bibnamefont
  {Dankert}}, \bibinfo {author} {\bibfnamefont {R.}~\bibnamefont {Cleve}},
  \bibinfo {author} {\bibfnamefont {J.}~\bibnamefont {Emerson}}, \ and\
  \bibinfo {author} {\bibfnamefont {E.}~\bibnamefont {Livine}},\ }\href@noop {}
  {\bibfield  {journal} {\bibinfo  {journal} {Phys. Rev. A}\ }\textbf {\bibinfo
  {volume} {80}},\ \bibinfo {pages} {012304} (\bibinfo {year}
  {2009}{\natexlab{a}})}\BibitemShut {NoStop}%
\bibitem [{\citenamefont {van Enk}\ and\ \citenamefont
  {Beenakker}(2012)}]{EB12}%
  \BibitemOpen
  \bibfield  {author} {\bibinfo {author} {\bibfnamefont {S.~J.}\ \bibnamefont
  {van Enk}}\ and\ \bibinfo {author} {\bibfnamefont {C.~W.~J.}\ \bibnamefont
  {Beenakker}},\ }\href@noop {} {\bibfield  {journal} {\bibinfo  {journal}
  {Phys. Rev. Lett.}\ }\textbf {\bibinfo {volume} {108}},\ \bibinfo {pages}
  {110503} (\bibinfo {year} {2012})}\BibitemShut {NoStop}%
\bibitem [{\citenamefont {Emerson}\ \emph {et~al.}(2005)\citenamefont
  {Emerson}, \citenamefont {Alicki},\ and\ \citenamefont
  {\.Zyczkowski}}]{Emerson05}%
  \BibitemOpen
  \bibfield  {author} {\bibinfo {author} {\bibfnamefont {J.}~\bibnamefont
  {Emerson}}, \bibinfo {author} {\bibfnamefont {R.}~\bibnamefont {Alicki}}, \
  and\ \bibinfo {author} {\bibfnamefont {K.}~\bibnamefont {\.Zyczkowski}},\
  }\href@noop {} {\bibfield  {journal} {\bibinfo  {journal} {J. Opt. B}\
  }\textbf {\bibinfo {volume} {7}},\ \bibinfo {pages} {S347} (\bibinfo {year}
  {2005})}\BibitemShut {NoStop}%
\bibitem [{\citenamefont {Magesan}\ \emph {et~al.}(2011)\citenamefont
  {Magesan}, \citenamefont {Gambetta},\ and\ \citenamefont {Emerson}}]{MGE11}%
  \BibitemOpen
  \bibfield  {author} {\bibinfo {author} {\bibfnamefont {E.}~\bibnamefont
  {Magesan}}, \bibinfo {author} {\bibfnamefont {J.~M.}\ \bibnamefont
  {Gambetta}}, \ and\ \bibinfo {author} {\bibfnamefont {J.}~\bibnamefont
  {Emerson}},\ }\href@noop {} {\bibfield  {journal} {\bibinfo  {journal} {Phys.
  Rev. Lett.}\ }\textbf {\bibinfo {volume} {106}},\ \bibinfo {pages} {180504}
  (\bibinfo {year} {2011})}\BibitemShut {NoStop}%
\bibitem [{\citenamefont {Roberts}\ and\ \citenamefont
  {Yoshida}(2017{\natexlab{a}})}]{RY17}%
  \BibitemOpen
  \bibfield  {author} {\bibinfo {author} {\bibfnamefont {D.~A.}\ \bibnamefont
  {Roberts}}\ and\ \bibinfo {author} {\bibfnamefont {B.}~\bibnamefont
  {Yoshida}},\ }\href@noop {} {\bibfield  {journal} {\bibinfo  {journal} {J.
  High Energy Phys.}\ }\textbf {\bibinfo {volume} {04}},\ \bibinfo {pages}
  {121} (\bibinfo {year} {2017}{\natexlab{a}})}\BibitemShut {NoStop}%
\bibitem [{\citenamefont {Cotler}\ \emph {et~al.}(2017)\citenamefont {Cotler},
  \citenamefont {Hunter-Jones}, \citenamefont {Liu},\ and\ \citenamefont
  {Yoshida}}]{CJLY17}%
  \BibitemOpen
  \bibfield  {author} {\bibinfo {author} {\bibfnamefont {J.}~\bibnamefont
  {Cotler}}, \bibinfo {author} {\bibfnamefont {N.}~\bibnamefont
  {Hunter-Jones}}, \bibinfo {author} {\bibfnamefont {J.}~\bibnamefont {Liu}}, \
  and\ \bibinfo {author} {\bibfnamefont {B.}~\bibnamefont {Yoshida}},\
  }\href@noop {} {\bibfield  {journal} {\bibinfo  {journal} {J. High Energy
  Phys.}\ }\textbf {\bibinfo {volume} {11}},\ \bibinfo {pages} {48} (\bibinfo
  {year} {2017})}\BibitemShut {NoStop}%
\bibitem [{\citenamefont {Kos}\ \emph {et~al.}(2018)\citenamefont {Kos},
  \citenamefont {Ljubotina},\ and\ \citenamefont {Prosen}}]{KLP18}%
  \BibitemOpen
  \bibfield  {author} {\bibinfo {author} {\bibfnamefont {P.}~\bibnamefont
  {Kos}}, \bibinfo {author} {\bibfnamefont {M.}~\bibnamefont {Ljubotina}}, \
  and\ \bibinfo {author} {\bibfnamefont {T.}~\bibnamefont {Prosen}},\
  }\href@noop {} {\bibfield  {journal} {\bibinfo  {journal} {Phys. Rev. X}\
  }\textbf {\bibinfo {volume} {8}},\ \bibinfo {pages} {021062} (\bibinfo {year}
  {2018})}\BibitemShut {NoStop}%
\bibitem [{\citenamefont {Oszmaniec}\ \emph {et~al.}(2016)\citenamefont
  {Oszmaniec}, \citenamefont {Augusiak}, \citenamefont {Gogolin}, \citenamefont
  {Ko\l{}ody\'{n}ski}, \citenamefont {Ac\'{i}n},\ and\ \citenamefont
  {Lewenstein}}]{OAGKAL16}%
  \BibitemOpen
  \bibfield  {author} {\bibinfo {author} {\bibfnamefont {M.}~\bibnamefont
  {Oszmaniec}}, \bibinfo {author} {\bibfnamefont {R.}~\bibnamefont {Augusiak}},
  \bibinfo {author} {\bibfnamefont {C.}~\bibnamefont {Gogolin}}, \bibinfo
  {author} {\bibfnamefont {J.}~\bibnamefont {Ko\l{}ody\'{n}ski}}, \bibinfo
  {author} {\bibfnamefont {A.}~\bibnamefont {Ac\'{i}n}}, \ and\ \bibinfo
  {author} {\bibfnamefont {M.}~\bibnamefont {Lewenstein}},\ }\href@noop {}
  {\bibfield  {journal} {\bibinfo  {journal} {Phys. Rev. X}\ }\textbf {\bibinfo
  {volume} {6}},\ \bibinfo {pages} {041044} (\bibinfo {year}
  {2016})}\BibitemShut {NoStop}%
\bibitem [{\citenamefont {Pozniak}\ \emph {et~al.}(1998)\citenamefont
  {Pozniak}, \citenamefont {Zyczkowski},\ and\ \citenamefont {Kus}}]{PZK98}%
  \BibitemOpen
  \bibfield  {author} {\bibinfo {author} {\bibfnamefont {M.}~\bibnamefont
  {Pozniak}}, \bibinfo {author} {\bibfnamefont {K.}~\bibnamefont {Zyczkowski}},
  \ and\ \bibinfo {author} {\bibfnamefont {M.}~\bibnamefont {Kus}},\
  }\href@noop {} {\bibfield  {journal} {\bibinfo  {journal} {J. Phys. A}\
  }\textbf {\bibinfo {volume} {31}},\ \bibinfo {pages} {1059} (\bibinfo {year}
  {1998})}\BibitemShut {NoStop}%
\bibitem [{\citenamefont {Dankert}\ \emph
  {et~al.}(2009{\natexlab{b}})\citenamefont {Dankert}, \citenamefont {Cleve},
  \citenamefont {Emerson},\ and\ \citenamefont {Livine}}]{Dankert09}%
  \BibitemOpen
  \bibfield  {author} {\bibinfo {author} {\bibfnamefont {C.}~\bibnamefont
  {Dankert}}, \bibinfo {author} {\bibfnamefont {R.}~\bibnamefont {Cleve}},
  \bibinfo {author} {\bibfnamefont {J.}~\bibnamefont {Emerson}}, \ and\
  \bibinfo {author} {\bibfnamefont {E.}~\bibnamefont {Livine}},\ }\href@noop {}
  {\bibfield  {journal} {\bibinfo  {journal} {Phys. Rev. A}\ }\textbf {\bibinfo
  {volume} {80}},\ \bibinfo {pages} {012304} (\bibinfo {year}
  {2009}{\natexlab{b}})}\BibitemShut {NoStop}%
\bibitem [{\citenamefont {DiVincenzo}\ \emph {et~al.}(2002)\citenamefont
  {DiVincenzo}, \citenamefont {Leung},\ and\ \citenamefont {Terhal}}]{DLT02}%
  \BibitemOpen
  \bibfield  {author} {\bibinfo {author} {\bibfnamefont {D.~P.}\ \bibnamefont
  {DiVincenzo}}, \bibinfo {author} {\bibfnamefont {D.~W.}\ \bibnamefont
  {Leung}}, \ and\ \bibinfo {author} {\bibfnamefont {B.~M.}\ \bibnamefont
  {Terhal}},\ }\href@noop {} {\bibfield  {journal} {\bibinfo  {journal} {IEEE
  Trans. Inf. Theory}\ }\textbf {\bibinfo {volume} {48}},\ \bibinfo {pages}
  {580} (\bibinfo {year} {2002})}\BibitemShut {NoStop}%
\bibitem [{\citenamefont {Emerson}\ \emph {et~al.}(2003)\citenamefont
  {Emerson}, \citenamefont {Weinstein}, \citenamefont {Saraceno}, \citenamefont
  {Lloyd},\ and\ \citenamefont {Cory}}]{Emerson03}%
  \BibitemOpen
  \bibfield  {author} {\bibinfo {author} {\bibfnamefont {J.}~\bibnamefont
  {Emerson}}, \bibinfo {author} {\bibfnamefont {Y.~S.}\ \bibnamefont
  {Weinstein}}, \bibinfo {author} {\bibfnamefont {M.}~\bibnamefont {Saraceno}},
  \bibinfo {author} {\bibfnamefont {S.}~\bibnamefont {Lloyd}}, \ and\ \bibinfo
  {author} {\bibfnamefont {D.~G.}\ \bibnamefont {Cory}},\ }\href@noop {}
  {\bibfield  {journal} {\bibinfo  {journal} {Science}\ }\textbf {\bibinfo
  {volume} {302}},\ \bibinfo {pages} {2098} (\bibinfo {year}
  {2003})}\BibitemShut {NoStop}%
\bibitem [{\citenamefont {Arnaud}\ and\ \citenamefont
  {Braun}(2008)}]{Arnaud08}%
  \BibitemOpen
  \bibfield  {author} {\bibinfo {author} {\bibfnamefont {L.}~\bibnamefont
  {Arnaud}}\ and\ \bibinfo {author} {\bibfnamefont {D.}~\bibnamefont {Braun}},\
  }\href@noop {} {\bibfield  {journal} {\bibinfo  {journal} {Phys. Rev. A}\
  }\textbf {\bibinfo {volume} {78}},\ \bibinfo {pages} {062329} (\bibinfo
  {year} {2008})}\BibitemShut {NoStop}%
\bibitem [{\citenamefont {Harrow}\ and\ \citenamefont {Low}(2009)}]{Harrow09}%
  \BibitemOpen
  \bibfield  {author} {\bibinfo {author} {\bibfnamefont {A.}~\bibnamefont
  {Harrow}}\ and\ \bibinfo {author} {\bibfnamefont {R.~A.}\ \bibnamefont
  {Low}},\ }\href@noop {} {\bibfield  {journal} {\bibinfo  {journal} {Commun.
  Math. Phys.}\ }\textbf {\bibinfo {volume} {291}},\ \bibinfo {pages} {257}
  (\bibinfo {year} {2009})}\BibitemShut {NoStop}%
\bibitem [{\citenamefont {Brown}\ and\ \citenamefont {Viola}(2010)}]{BV10}%
  \BibitemOpen
  \bibfield  {author} {\bibinfo {author} {\bibfnamefont {W.~G.}\ \bibnamefont
  {Brown}}\ and\ \bibinfo {author} {\bibfnamefont {L.}~\bibnamefont {Viola}},\
  }\href@noop {} {\bibfield  {journal} {\bibinfo  {journal} {Phys. Rev. Lett.}\
  }\textbf {\bibinfo {volume} {104}},\ \bibinfo {pages} {250501} (\bibinfo
  {year} {2010})}\BibitemShut {NoStop}%
\bibitem [{\citenamefont {\'{C}wikli\'{n}ski}\ \emph
  {et~al.}(2013)\citenamefont {\'{C}wikli\'{n}ski}, \citenamefont {Horodecki},
  \citenamefont {Mozrzymas}, \citenamefont {Pankowski},\ and\ \citenamefont
  {Studzi\'{n}ski}}]{CHMPS13}%
  \BibitemOpen
  \bibfield  {author} {\bibinfo {author} {\bibfnamefont {P.}~\bibnamefont
  {\'{C}wikli\'{n}ski}}, \bibinfo {author} {\bibfnamefont {M.}~\bibnamefont
  {Horodecki}}, \bibinfo {author} {\bibfnamefont {M.}~\bibnamefont
  {Mozrzymas}}, \bibinfo {author} {\bibfnamefont {{\L}.}~\bibnamefont
  {Pankowski}}, \ and\ \bibinfo {author} {\bibfnamefont {M.}~\bibnamefont
  {Studzi\'{n}ski}},\ }\href@noop {} {\bibfield  {journal} {\bibinfo  {journal}
  {J. Phys. A: Math. Theor.}\ }\textbf {\bibinfo {volume} {46}},\ \bibinfo
  {pages} {305301} (\bibinfo {year} {2013})}\BibitemShut {NoStop}%
\bibitem [{\citenamefont {Brand{\~a}o}\ \emph {et~al.}(2016)\citenamefont
  {Brand{\~a}o}, \citenamefont {Harrow},\ and\ \citenamefont
  {Horodecki}}]{brandao2016efficient}%
  \BibitemOpen
  \bibfield  {author} {\bibinfo {author} {\bibfnamefont {F.~G.}\ \bibnamefont
  {Brand{\~a}o}}, \bibinfo {author} {\bibfnamefont {A.~W.}\ \bibnamefont
  {Harrow}}, \ and\ \bibinfo {author} {\bibfnamefont {M.}~\bibnamefont
  {Horodecki}},\ }\href@noop {} {\bibfield  {journal} {\bibinfo  {journal}
  {Phys. Rev. Lett.}\ }\textbf {\bibinfo {volume} {116}},\ \bibinfo {pages}
  {170502} (\bibinfo {year} {2016})}\BibitemShut {NoStop}%
\bibitem [{\citenamefont {Emerson}\ \emph {et~al.}(2007)\citenamefont
  {Emerson}, \citenamefont {Silva}, \citenamefont {Moussa}, \citenamefont
  {Ryan}, \citenamefont {Laforest}, \citenamefont {Baugh}, \citenamefont
  {Cory},\ and\ \citenamefont {Laflamme}}]{Emerson07}%
  \BibitemOpen
  \bibfield  {author} {\bibinfo {author} {\bibfnamefont {J.}~\bibnamefont
  {Emerson}}, \bibinfo {author} {\bibfnamefont {M.}~\bibnamefont {Silva}},
  \bibinfo {author} {\bibfnamefont {O.}~\bibnamefont {Moussa}}, \bibinfo
  {author} {\bibfnamefont {C.}~\bibnamefont {Ryan}}, \bibinfo {author}
  {\bibfnamefont {M.}~\bibnamefont {Laforest}}, \bibinfo {author}
  {\bibfnamefont {J.}~\bibnamefont {Baugh}}, \bibinfo {author} {\bibfnamefont
  {D.~G.}\ \bibnamefont {Cory}}, \ and\ \bibinfo {author} {\bibfnamefont
  {R.}~\bibnamefont {Laflamme}},\ }\href@noop {} {\bibfield  {journal}
  {\bibinfo  {journal} {Science}\ }\textbf {\bibinfo {volume} {317}},\ \bibinfo
  {pages} {1893} (\bibinfo {year} {2007})}\BibitemShut {NoStop}%
\bibitem [{\citenamefont {Chow}\ \emph {et~al.}(2009)\citenamefont {Chow},
  \citenamefont {Gambetta}, \citenamefont {Tornberg}, \citenamefont {Koch},
  \citenamefont {Bishop}, \citenamefont {Houck}, \citenamefont {Johnson},
  \citenamefont {Frunzio}, \citenamefont {Girvin},\ and\ \citenamefont
  {Schoelkopf}}]{Chow09}%
  \BibitemOpen
  \bibfield  {author} {\bibinfo {author} {\bibfnamefont {J.~M.}\ \bibnamefont
  {Chow}}, \bibinfo {author} {\bibfnamefont {J.~M.}\ \bibnamefont {Gambetta}},
  \bibinfo {author} {\bibfnamefont {L.}~\bibnamefont {Tornberg}}, \bibinfo
  {author} {\bibfnamefont {J.}~\bibnamefont {Koch}}, \bibinfo {author}
  {\bibfnamefont {L.~S.}\ \bibnamefont {Bishop}}, \bibinfo {author}
  {\bibfnamefont {A.~A.}\ \bibnamefont {Houck}}, \bibinfo {author}
  {\bibfnamefont {B.~R.}\ \bibnamefont {Johnson}}, \bibinfo {author}
  {\bibfnamefont {L.}~\bibnamefont {Frunzio}}, \bibinfo {author} {\bibfnamefont
  {S.~M.}\ \bibnamefont {Girvin}}, \ and\ \bibinfo {author} {\bibfnamefont
  {R.~J.}\ \bibnamefont {Schoelkopf}},\ }\href@noop {} {\bibfield  {journal}
  {\bibinfo  {journal} {Phys. Rev. Lett.}\ }\textbf {\bibinfo {volume} {102}},\
  \bibinfo {pages} {090502} (\bibinfo {year} {2009})}\BibitemShut {NoStop}%
\bibitem [{\citenamefont {C\'{o}rcoles}\ \emph {et~al.}(2013)\citenamefont
  {C\'{o}rcoles}, \citenamefont {Gambetta}, \citenamefont {Chow}, \citenamefont
  {Smolin}, \citenamefont {Ware}, \citenamefont {Strand}, \citenamefont
  {Plourde},\ and\ \citenamefont {Steffen}}]{Steffen13}%
  \BibitemOpen
  \bibfield  {author} {\bibinfo {author} {\bibfnamefont {A.~D.}\ \bibnamefont
  {C\'{o}rcoles}}, \bibinfo {author} {\bibfnamefont {J.~M.}\ \bibnamefont
  {Gambetta}}, \bibinfo {author} {\bibfnamefont {J.~M.}\ \bibnamefont {Chow}},
  \bibinfo {author} {\bibfnamefont {J.~A.}\ \bibnamefont {Smolin}}, \bibinfo
  {author} {\bibfnamefont {M.}~\bibnamefont {Ware}}, \bibinfo {author}
  {\bibfnamefont {J.}~\bibnamefont {Strand}}, \bibinfo {author} {\bibfnamefont
  {B.~L.~T.}\ \bibnamefont {Plourde}}, \ and\ \bibinfo {author} {\bibfnamefont
  {M.}~\bibnamefont {Steffen}},\ }\href@noop {} {\bibfield  {journal} {\bibinfo
   {journal} {Phys. Rev. A}\ }\textbf {\bibinfo {volume} {87}},\ \bibinfo
  {pages} {030301} (\bibinfo {year} {2013})}\BibitemShut {NoStop}%
\bibitem [{\citenamefont {Matthews}\ \emph {et~al.}(2015)\citenamefont
  {Matthews}, \citenamefont {Whittaker}, \citenamefont {O'Brien},\ and\
  \citenamefont {Turner}}]{MWOT15}%
  \BibitemOpen
  \bibfield  {author} {\bibinfo {author} {\bibfnamefont {J.~C.~F.}\
  \bibnamefont {Matthews}}, \bibinfo {author} {\bibfnamefont {R.}~\bibnamefont
  {Whittaker}}, \bibinfo {author} {\bibfnamefont {J.~L.}\ \bibnamefont
  {O'Brien}}, \ and\ \bibinfo {author} {\bibfnamefont {P.~S.}\ \bibnamefont
  {Turner}},\ }\href@noop {} {\bibfield  {journal} {\bibinfo  {journal} {Phys.
  Rev. A}\ }\textbf {\bibinfo {volume} {91}},\ \bibinfo {pages} {020301}
  (\bibinfo {year} {2015})}\BibitemShut {NoStop}%
\bibitem [{\citenamefont {Turner}\ and\ \citenamefont {Markham}(2016)}]{TM16}%
  \BibitemOpen
  \bibfield  {author} {\bibinfo {author} {\bibfnamefont {P.~S.}\ \bibnamefont
  {Turner}}\ and\ \bibinfo {author} {\bibfnamefont {D.}~\bibnamefont
  {Markham}},\ }\href@noop {} {\bibfield  {journal} {\bibinfo  {journal} {Phys.
  Rev. Lett.}\ }\textbf {\bibinfo {volume} {116}},\ \bibinfo {pages} {200501}
  (\bibinfo {year} {2016})}\BibitemShut {NoStop}%
\bibitem [{\citenamefont {Mezher}\ \emph {et~al.}(2018)\citenamefont {Mezher},
  \citenamefont {Ghalbouni}, \citenamefont {Dgheim},\ and\ \citenamefont
  {Markham}}]{MGDM18}%
  \BibitemOpen
  \bibfield  {author} {\bibinfo {author} {\bibfnamefont {R.}~\bibnamefont
  {Mezher}}, \bibinfo {author} {\bibfnamefont {J.}~\bibnamefont {Ghalbouni}},
  \bibinfo {author} {\bibfnamefont {J.}~\bibnamefont {Dgheim}}, \ and\ \bibinfo
  {author} {\bibfnamefont {D.}~\bibnamefont {Markham}},\ }\href@noop {}
  {\bibfield  {journal} {\bibinfo  {journal} {Phys. Rev. A}\ }\textbf {\bibinfo
  {volume} {97}},\ \bibinfo {pages} {022333} (\bibinfo {year}
  {2018})}\BibitemShut {NoStop}%
\bibitem [{\citenamefont {Nakata}\ \emph {et~al.}(2017)\citenamefont {Nakata},
  \citenamefont {Hirche}, \citenamefont {Koashi},\ and\ \citenamefont
  {Winter}}]{nakata2017efficient}%
  \BibitemOpen
  \bibfield  {author} {\bibinfo {author} {\bibfnamefont {Y.}~\bibnamefont
  {Nakata}}, \bibinfo {author} {\bibfnamefont {C.}~\bibnamefont {Hirche}},
  \bibinfo {author} {\bibfnamefont {M.}~\bibnamefont {Koashi}}, \ and\ \bibinfo
  {author} {\bibfnamefont {A.}~\bibnamefont {Winter}},\ }\href@noop {}
  {\bibfield  {journal} {\bibinfo  {journal} {Phys. Rev. X}\ }\textbf {\bibinfo
  {volume} {7}},\ \bibinfo {pages} {021006} (\bibinfo {year}
  {2017})}\BibitemShut {NoStop}%
\bibitem [{\citenamefont {Roberts}\ and\ \citenamefont
  {Yoshida}(2017{\natexlab{b}})}]{Yoshida2017}%
  \BibitemOpen
  \bibfield  {author} {\bibinfo {author} {\bibfnamefont {D.~A.}\ \bibnamefont
  {Roberts}}\ and\ \bibinfo {author} {\bibfnamefont {B.}~\bibnamefont
  {Yoshida}},\ }\href@noop {} {\bibfield  {journal} {\bibinfo  {journal} {J.
  High Energy Phys.}\ }\textbf {\bibinfo {volume} {04}},\ \bibinfo {pages}
  {121} (\bibinfo {year} {2017}{\natexlab{b}})}\BibitemShut {NoStop}%
\bibitem [{\citenamefont {Liu}\ \emph {et~al.}(2018)\citenamefont {Liu},
  \citenamefont {Lloyd}, \citenamefont {Zhu},\ and\ \citenamefont
  {Zhu}}]{LLZZ18}%
  \BibitemOpen
  \bibfield  {author} {\bibinfo {author} {\bibfnamefont {Z.-W.}\ \bibnamefont
  {Liu}}, \bibinfo {author} {\bibfnamefont {S.}~\bibnamefont {Lloyd}}, \bibinfo
  {author} {\bibfnamefont {E.~Y.}\ \bibnamefont {Zhu}}, \ and\ \bibinfo
  {author} {\bibfnamefont {H.}~\bibnamefont {Zhu}},\ }\href@noop {} {\bibfield
  {journal} {\bibinfo  {journal} {Phys. Rev. Lett.}\ }\textbf {\bibinfo
  {volume} {120}},\ \bibinfo {pages} {130502} (\bibinfo {year}
  {2018})}\BibitemShut {NoStop}%
\bibitem [{\citenamefont {Alves}\ and\ \citenamefont {Flynn}(2018)}]{AF18}%
  \BibitemOpen
  \bibfield  {author} {\bibinfo {author} {\bibfnamefont {D.~W.~F.}\
  \bibnamefont {Alves}}\ and\ \bibinfo {author} {\bibfnamefont {M.~O.}\
  \bibnamefont {Flynn}},\ }\href@noop {} {\bibfield  {journal} {\bibinfo
  {journal} {arXiv:1808.10498v1}\ } (\bibinfo {year} {2018})}\BibitemShut
  {NoStop}%
\bibitem [{\citenamefont {Baum}\ \emph {et~al.}(1985)\citenamefont {Baum},
  \citenamefont {Munowitz}, \citenamefont {Garroway},\ and\ \citenamefont
  {Pines}}]{Baum85}%
  \BibitemOpen
  \bibfield  {author} {\bibinfo {author} {\bibfnamefont {J.}~\bibnamefont
  {Baum}}, \bibinfo {author} {\bibfnamefont {M.}~\bibnamefont {Munowitz}},
  \bibinfo {author} {\bibfnamefont {A.~N.}\ \bibnamefont {Garroway}}, \ and\
  \bibinfo {author} {\bibfnamefont {A.}~\bibnamefont {Pines}},\ }\href@noop {}
  {\bibfield  {journal} {\bibinfo  {journal} {J. Chem. Phys.}\ }\textbf
  {\bibinfo {volume} {83}},\ \bibinfo {pages} {2015} (\bibinfo {year}
  {1985})}\BibitemShut {NoStop}%
\bibitem [{\citenamefont {Lacelle}(1991)}]{Lacelle91}%
  \BibitemOpen
  \bibfield  {author} {\bibinfo {author} {\bibfnamefont {S.}~\bibnamefont
  {Lacelle}},\ }\href@noop {} {\bibfield  {journal} {\bibinfo  {journal} {Adv.
  Magn. Opt. Reson.}\ }\textbf {\bibinfo {volume} {16}},\ \bibinfo {pages}
  {173} (\bibinfo {year} {1991})}\BibitemShut {NoStop}%
\bibitem [{\citenamefont {\'{A}lvarez}\ and\ \citenamefont
  {Suter}(2010)}]{AS10}%
  \BibitemOpen
  \bibfield  {author} {\bibinfo {author} {\bibfnamefont {G.~A.}\ \bibnamefont
  {\'{A}lvarez}}\ and\ \bibinfo {author} {\bibfnamefont {D.}~\bibnamefont
  {Suter}},\ }\href@noop {} {\bibfield  {journal} {\bibinfo  {journal} {Phys.
  Rev. Lett.}\ }\textbf {\bibinfo {volume} {104}},\ \bibinfo {pages} {230403}
  (\bibinfo {year} {2010})}\BibitemShut {NoStop}%
\bibitem [{\citenamefont {\'{A}lvarez}\ \emph {et~al.}(2015)\citenamefont
  {\'{A}lvarez}, \citenamefont {Suter},\ and\ \citenamefont {Kaiser}}]{ASK15}%
  \BibitemOpen
  \bibfield  {author} {\bibinfo {author} {\bibfnamefont {G.~A.}\ \bibnamefont
  {\'{A}lvarez}}, \bibinfo {author} {\bibfnamefont {D.}~\bibnamefont {Suter}},
  \ and\ \bibinfo {author} {\bibfnamefont {R.}~\bibnamefont {Kaiser}},\
  }\href@noop {} {\bibfield  {journal} {\bibinfo  {journal} {Science}\ }\textbf
  {\bibinfo {volume} {349}},\ \bibinfo {pages} {846} (\bibinfo {year}
  {2015})}\BibitemShut {NoStop}%
\bibitem [{\citenamefont {Wei}\ \emph {et~al.}(2018)\citenamefont {Wei},
  \citenamefont {Ramanathan},\ and\ \citenamefont {Cappellaro}}]{WRC18}%
  \BibitemOpen
  \bibfield  {author} {\bibinfo {author} {\bibfnamefont {K.~X.}\ \bibnamefont
  {Wei}}, \bibinfo {author} {\bibfnamefont {C.}~\bibnamefont {Ramanathan}}, \
  and\ \bibinfo {author} {\bibfnamefont {P.}~\bibnamefont {Cappellaro}},\
  }\href@noop {} {\bibfield  {journal} {\bibinfo  {journal} {Phys. Rev. Lett.}\
  }\textbf {\bibinfo {volume} {120}},\ \bibinfo {pages} {070501} (\bibinfo
  {year} {2018})}\BibitemShut {NoStop}%
\bibitem [{\citenamefont {G\"{a}rttner}\ \emph {et~al.}(2018)\citenamefont
  {G\"{a}rttner}, \citenamefont {Hauke},\ and\ \citenamefont {Rey}}]{GHR18}%
  \BibitemOpen
  \bibfield  {author} {\bibinfo {author} {\bibfnamefont {M.}~\bibnamefont
  {G\"{a}rttner}}, \bibinfo {author} {\bibfnamefont {P.}~\bibnamefont {Hauke}},
  \ and\ \bibinfo {author} {\bibfnamefont {A.~M.}\ \bibnamefont {Rey}},\
  }\href@noop {} {\bibfield  {journal} {\bibinfo  {journal} {Phys. Rev. Lett.}\
  }\textbf {\bibinfo {volume} {120}},\ \bibinfo {pages} {040402} (\bibinfo
  {year} {2018})}\BibitemShut {NoStop}%
\bibitem [{\citenamefont {G\"{a}rttner}\ \emph {et~al.}(2017)\citenamefont
  {G\"{a}rttner}, \citenamefont {Bohnet}, \citenamefont {Safavi-Naini},
  \citenamefont {Wall}, \citenamefont {Bollinger},\ and\ \citenamefont
  {Rey}}]{Garttner17}%
  \BibitemOpen
  \bibfield  {author} {\bibinfo {author} {\bibfnamefont {M.}~\bibnamefont
  {G\"{a}rttner}}, \bibinfo {author} {\bibfnamefont {J.~G.}\ \bibnamefont
  {Bohnet}}, \bibinfo {author} {\bibfnamefont {A.}~\bibnamefont
  {Safavi-Naini}}, \bibinfo {author} {\bibfnamefont {M.~L.}\ \bibnamefont
  {Wall}}, \bibinfo {author} {\bibfnamefont {J.~J.}\ \bibnamefont {Bollinger}},
  \ and\ \bibinfo {author} {\bibfnamefont {A.~M.}\ \bibnamefont {Rey}},\
  }\href@noop {} {\bibfield  {journal} {\bibinfo  {journal} {Nat. Phys.}\
  }\textbf {\bibinfo {volume} {13}},\ \bibinfo {pages} {781} (\bibinfo {year}
  {2017})}\BibitemShut {NoStop}%
\bibitem [{S()}]{S}%
  \BibitemOpen
  \href@noop {} {\ }\bibinfo {note} {See Supplementary Material for more
  details}\BibitemShut {NoStop}%
\bibitem [{\citenamefont {Vandersypen}\ and\ \citenamefont
  {Chuang}(2004)}]{VC04}%
  \BibitemOpen
  \bibfield  {author} {\bibinfo {author} {\bibfnamefont {L.~M.~K.}\
  \bibnamefont {Vandersypen}}\ and\ \bibinfo {author} {\bibfnamefont {I.~L.}\
  \bibnamefont {Chuang}},\ }\href@noop {} {\bibfield  {journal} {\bibinfo
  {journal} {Rev. Mod. Phys.}\ }\textbf {\bibinfo {volume} {76}},\ \bibinfo
  {pages} {1037} (\bibinfo {year} {2004})}\BibitemShut {NoStop}%
\bibitem [{\citenamefont {Santos}\ and\ \citenamefont {Viola}(2006)}]{SV06}%
  \BibitemOpen
  \bibfield  {author} {\bibinfo {author} {\bibfnamefont {L.~F.}\ \bibnamefont
  {Santos}}\ and\ \bibinfo {author} {\bibfnamefont {L.}~\bibnamefont {Viola}},\
  }\href@noop {} {\bibfield  {journal} {\bibinfo  {journal} {Phys. Rev. Lett.}\
  }\textbf {\bibinfo {volume} {97}},\ \bibinfo {pages} {150501} (\bibinfo
  {year} {2006})}\BibitemShut {NoStop}%
\bibitem [{\citenamefont {Santos}\ and\ \citenamefont {Viola}(2008)}]{SV08}%
  \BibitemOpen
  \bibfield  {author} {\bibinfo {author} {\bibfnamefont {L.~F.}\ \bibnamefont
  {Santos}}\ and\ \bibinfo {author} {\bibfnamefont {L.}~\bibnamefont {Viola}},\
  }\href@noop {} {\bibfield  {journal} {\bibinfo  {journal} {New J. Phys.}\
  }\textbf {\bibinfo {volume} {10}},\ \bibinfo {pages} {083009} (\bibinfo
  {year} {2008})}\BibitemShut {NoStop}%
\bibitem [{\citenamefont {Renes}\ \emph {et~al.}(2004)\citenamefont {Renes},
  \citenamefont {Blume-Kohout}, \citenamefont {Scott},\ and\ \citenamefont
  {Caves}}]{RBSC04}%
  \BibitemOpen
  \bibfield  {author} {\bibinfo {author} {\bibfnamefont {J.~M.}\ \bibnamefont
  {Renes}}, \bibinfo {author} {\bibfnamefont {R.}~\bibnamefont {Blume-Kohout}},
  \bibinfo {author} {\bibfnamefont {A.~J.}\ \bibnamefont {Scott}}, \ and\
  \bibinfo {author} {\bibfnamefont {C.~M.}\ \bibnamefont {Caves}},\ }\href@noop
  {} {\bibfield  {journal} {\bibinfo  {journal} {J. Math. Phys.}\ }\textbf
  {\bibinfo {volume} {45}},\ \bibinfo {pages} {2171} (\bibinfo {year}
  {2004})}\BibitemShut {NoStop}%
\bibitem [{\citenamefont {Scott}(2008)}]{Scott2008}%
  \BibitemOpen
  \bibfield  {author} {\bibinfo {author} {\bibfnamefont {A.~J.}\ \bibnamefont
  {Scott}},\ }\href@noop {} {\bibfield  {journal} {\bibinfo  {journal} {J.
  Phys. A}\ }\textbf {\bibinfo {volume} {41}},\ \bibinfo {pages} {055308}
  (\bibinfo {year} {2008})}\BibitemShut {NoStop}%
\bibitem [{\citenamefont {Roy}\ and\ \citenamefont {Scott}(2009)}]{RS09}%
  \BibitemOpen
  \bibfield  {author} {\bibinfo {author} {\bibfnamefont {A.}~\bibnamefont
  {Roy}}\ and\ \bibinfo {author} {\bibfnamefont {A.~J.}\ \bibnamefont
  {Scott}},\ }\href@noop {} {\bibfield  {journal} {\bibinfo  {journal} {Des.
  Code Cryptogr.}\ }\textbf {\bibinfo {volume} {53}},\ \bibinfo {pages} {13}
  (\bibinfo {year} {2009})}\BibitemShut {NoStop}%
\bibitem [{\citenamefont {Li}\ \emph {et~al.}(2017)\citenamefont {Li},
  \citenamefont {Fan}, \citenamefont {Wang}, \citenamefont {Ye},
  \citenamefont {Zeng}, \citenamefont {Zhai}, \citenamefont {Peng},\ and\
  \citenamefont {Du}}]{Jun17}%
  \BibitemOpen
  \bibfield  {author} {\bibinfo {author} {\bibfnamefont {J.}~\bibnamefont
  {Li}}, \bibinfo {author} {\bibfnamefont {R.}~\bibnamefont {Fan}}, \bibinfo
  {author} {\bibfnamefont {H.}~\bibnamefont {Wang}}, \bibinfo {author} {\bibfnamefont
  {B.}~\bibnamefont {Ye}}, \bibinfo {author} {\bibfnamefont {B.}~\bibnamefont
  {Zeng}}, \bibinfo {author} {\bibfnamefont {H.}~\bibnamefont {Zhai}}, \bibinfo
  {author} {\bibfnamefont {X.}~\bibnamefont {Peng}}, \ and\ \bibinfo {author}
  {\bibfnamefont {J.}~\bibnamefont {Du}},\ }\href@noop {} {\bibfield  {journal}
  {\bibinfo  {journal} {Phys. Rev. X}\ }\textbf {\bibinfo {volume} {7}},\
  \bibinfo {pages} {031011} (\bibinfo {year} {2017})}\BibitemShut {NoStop}%
\bibitem [{\citenamefont {Ryan}\ \emph {et~al.}(2008)\citenamefont {Ryan},
  \citenamefont {Negrevergne}, \citenamefont {Laforest}, \citenamefont
  {Knill},\ and\ \citenamefont {Laflamme}}]{Ryan08}%
  \BibitemOpen
  \bibfield  {author} {\bibinfo {author} {\bibfnamefont {C.~A.}\ \bibnamefont
  {Ryan}}, \bibinfo {author} {\bibfnamefont {C.}~\bibnamefont {Negrevergne}},
  \bibinfo {author} {\bibfnamefont {M.}~\bibnamefont {Laforest}}, \bibinfo
  {author} {\bibfnamefont {E.}~\bibnamefont {Knill}}, \ and\ \bibinfo {author}
  {\bibfnamefont {R.}~\bibnamefont {Laflamme}},\ }\href@noop {} {\bibfield
  {journal} {\bibinfo  {journal} {Science}\ }\textbf {\bibinfo {volume} {78}},\
  \bibinfo {pages} {012328} (\bibinfo {year} {2008})}\BibitemShut {NoStop}%
\bibitem [{\citenamefont {Li}\ \emph {et~al.}(2016)\citenamefont {Li},
  \citenamefont {Cui}, \citenamefont {Laflamme},\ and\ \citenamefont
  {Peng}}]{Jun16}%
  \BibitemOpen
  \bibfield  {author} {\bibinfo {author} {\bibfnamefont {J.}~\bibnamefont
  {Li}}, \bibinfo {author} {\bibfnamefont {J.}~\bibnamefont {Cui}}, \bibinfo
  {author} {\bibfnamefont {R.}~\bibnamefont {Laflamme}}, \ and\ \bibinfo
  {author} {\bibfnamefont {X.}~\bibnamefont {Peng}},\ }\href@noop {} {\bibfield
   {journal} {\bibinfo  {journal} {Phys. Rev. A}\ }\textbf {\bibinfo {volume}
  {94}},\ \bibinfo {pages} {032316} (\bibinfo {year} {2016})}\BibitemShut
  {NoStop}%
\end{thebibliography}
\end{document}